\documentclass[%
 aip,
 amsmath,amssymb,
 reprint,%
]{revtex4-1}

\usepackage{graphicx}
\usepackage{dcolumn}
\usepackage{bm}

\usepackage[utf8]{inputenc}
\usepackage[T1]{fontenc}
\usepackage{mathptmx}

\begin{document}

\preprint{AIP/123-QED}

\title[A High-Dynamic-Range Digital RF-Over-Fiber Link for MRI Receive Coils Using Delta-Sigma Modulation]{A High-Dynamic-Range Digital RF-Over-Fiber Link for MRI Receive Coils Using Delta-Sigma Modulation}

\author{Mingdong~Fan}%
\email{mxf293@case.edu}
\author{Robert~W.~Brown}%
\email{rwb@case.edu}
\affiliation{%
Physics Department, Case Western Reserve University, Cleveland, OH 44106, USA.
}
\author{Xi~Gao}%
 \email{xxg171@case.edu}
\author{Soumyajit~Mandal}
\altaffiliation[Author ]{to whom correspondence should be addressed}
\email{sxm833@case.edu.}
\affiliation{%
Electrical, Computer, and Systems Engineering Department, Case Western Reserve University, Cleveland, OH 44106, USA.
}%
\author{Labros~Petropoulos}%
\email{labros.petropoulos@gualedyn.com}
\author{Xiaoyu~Yang}%
\email{xiaoyu.yang@qualedyn.com}
\author{Shinya~Handa}%
\email{shinya.handa@qualedyn.com}
\author{Hiroyuki~Fujita}%
\email{hiroyuki.fujita@ qualedyn.com}
\affiliation{ 
Quality Electrodynamics, LLC, Mayfield, OH 44143, USA.
}%

\date{\today}

\begin{abstract}
The coaxial cables commonly used to connect RF coil arrays with the control console of an MRI scanner are susceptible to electromagnetic coupling. As the number of RF channel increases, such coupling could result in severe heating and pose a safety concern. Non-conductive transmission solutions based on fiber-optic cables are considered to be one of the alternatives, but are limited by the high dynamic range ($>80$~dB) of typical MRI signals. A new digital fiber-optic transmission system based on delta-sigma modulation (DSM) is developed to address this problem. A DSM-based optical link is prototyped using off-the-shelf components and bench-tested at different signal oversampling rates (OSR). An end-to-end dynamic range (DR) of 81~dB, which is sufficient for typical MRI signals, is obtained over a bandwidth of 200~kHz, which corresponds to $OSR=50$. A fully-integrated custom fourth-order continuous-time DSM (CT-DSM) is designed in 180~nm CMOS technology to enable transmission of full-bandwidth MRI signals (up to 1~MHz) with adequate DR. Initial electrical test results from this custom chip are also presented.
\end{abstract}

\maketitle

\section{Introduction}
\label{sec:intro}
Surface receive array coils are often used to detect the radio-frequency (RF) signals in MRI~\cite{ohliger2006introduction,fujita2013rf}. Because of their major advantages in i) image signal-to-noise ratio (SNR) over an extended field-of-view (FOV), and ii) imaging time, the sizes of such arrays (and the number of associated RF transceiver channels) have been increasing over the years, with as many as 128 channels now being manufactured for clinical applications~\cite{gruber2018rf}. The standard method to transmit the MRI signal from receive coil elements in such arrays to the control console is through a coaxial cable. However, as the number of RF coils increases, the electromagnetic coupling between coaxial cables becomes a severe problem. Such coupling can potentially lead to signal loss and unnecessary consumption of transmit power, affect the image quality, and also cause heating in nearby coils and cables that can be a major safety concern if the cables are very close to the patients~\cite{armenean2004rf, yeung2002rf}. Efforts have been made to estimate and reduce the coupling in coaxial cables~\cite{armenean2004rf, yeung2002rf, weiss2005transmission, peterson2003common, lee2002coupling, ladd2000reduction}, but the complexity of the shield current makes it very difficult to completely eliminate it all along the cable. In addition to the electromagnetic coupling, the bulkiness from the large number of cables and baluns contributes to physical difficulties such as space and weight.

Several transmission technologies have started to emerge as alternatives to coaxial cables. Wireless transmission of MRI signals has been investigated by several groups because it can greatly simplify the receive array setup by being completely cable-free~\cite{aggarwal2016millimeter, riffe2014wireless, riffe2013identification, heid2009cutting, wei2007realization, scott2005wireless}. However, there are still major challenges, one of which is the power consumption. Because the transmit power radiates into space, significant pass loss is unavoidable and only a small portion of the power can be received by the receive antenna~\cite{riffe2014wireless, riffe2013identification}. In addition, most proposed wireless systems rely on batteries to power the active components, and it is not feasible to use battery power to support multiple RF coils with such high power consumption~\cite{aggarwal2016millimeter, riffe2014wireless, riffe2013identification, wei2007realization}. 

Fiber-optical links are considered as another promising alternative to coaxial cables. Optical fibers are small and light, and most importantly are immune to electromagnetic coupling. The light is also confined within the fiber and does not geometrically spread as it propagates, thus reducing power requirements compared to wireless systems. Analog optical links have been studied due to their relatively simple RF coil structure and minimal modifications to the MRI system~\cite{memis2008miniaturized,biber2008analog,yuan2007direct,koste2005optical,du2007comparison,ayde2014unbiased,koste2009integrated,yuan2007investigation,yuan20084}, but are limited by electrical and optical nonlinearities that degrade noise figure and dynamic range (DR).

\begin{figure*}
    \centering
    \includegraphics[width=0.34\textwidth]{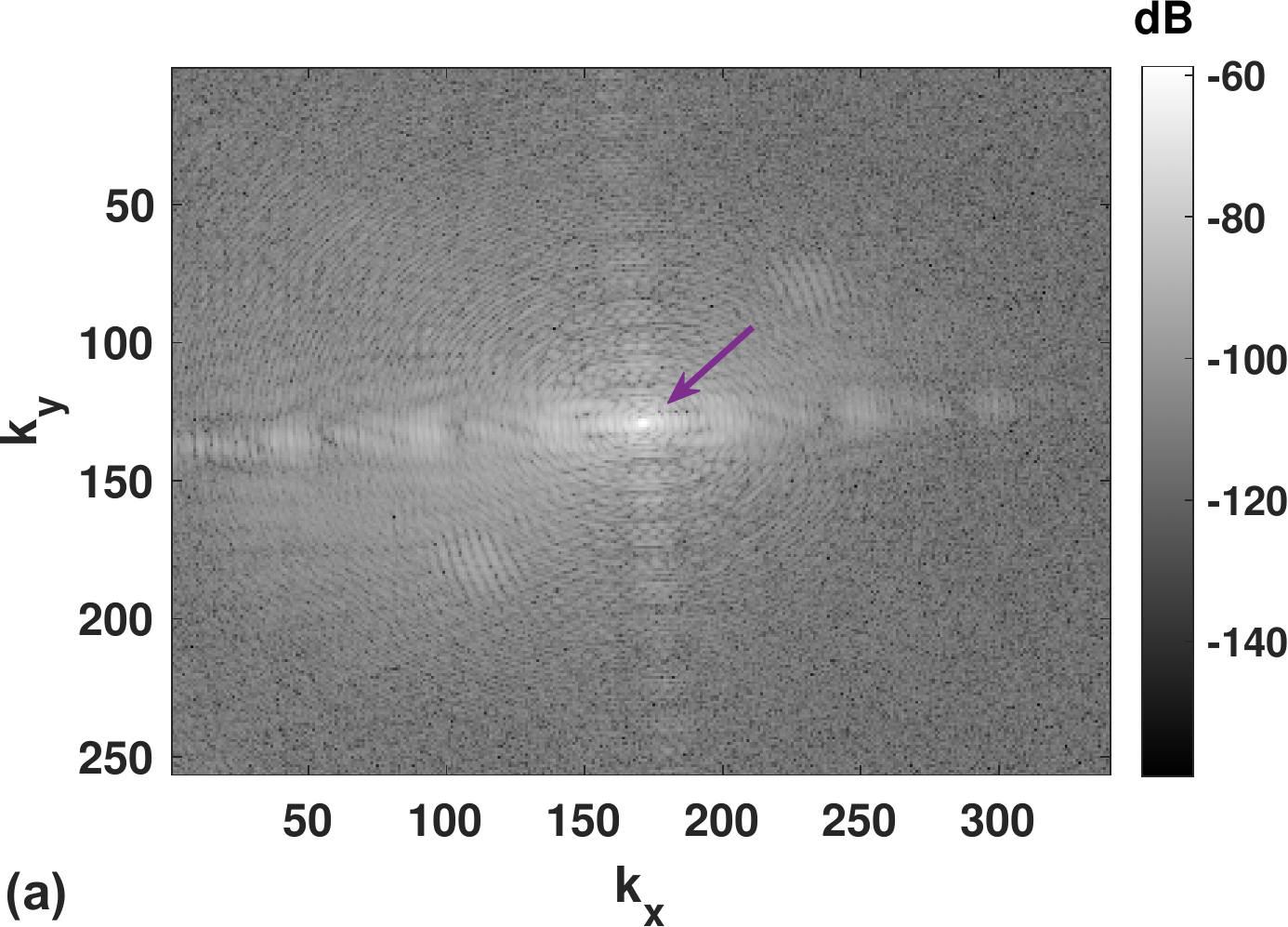}
    \includegraphics[width=0.28\textwidth]{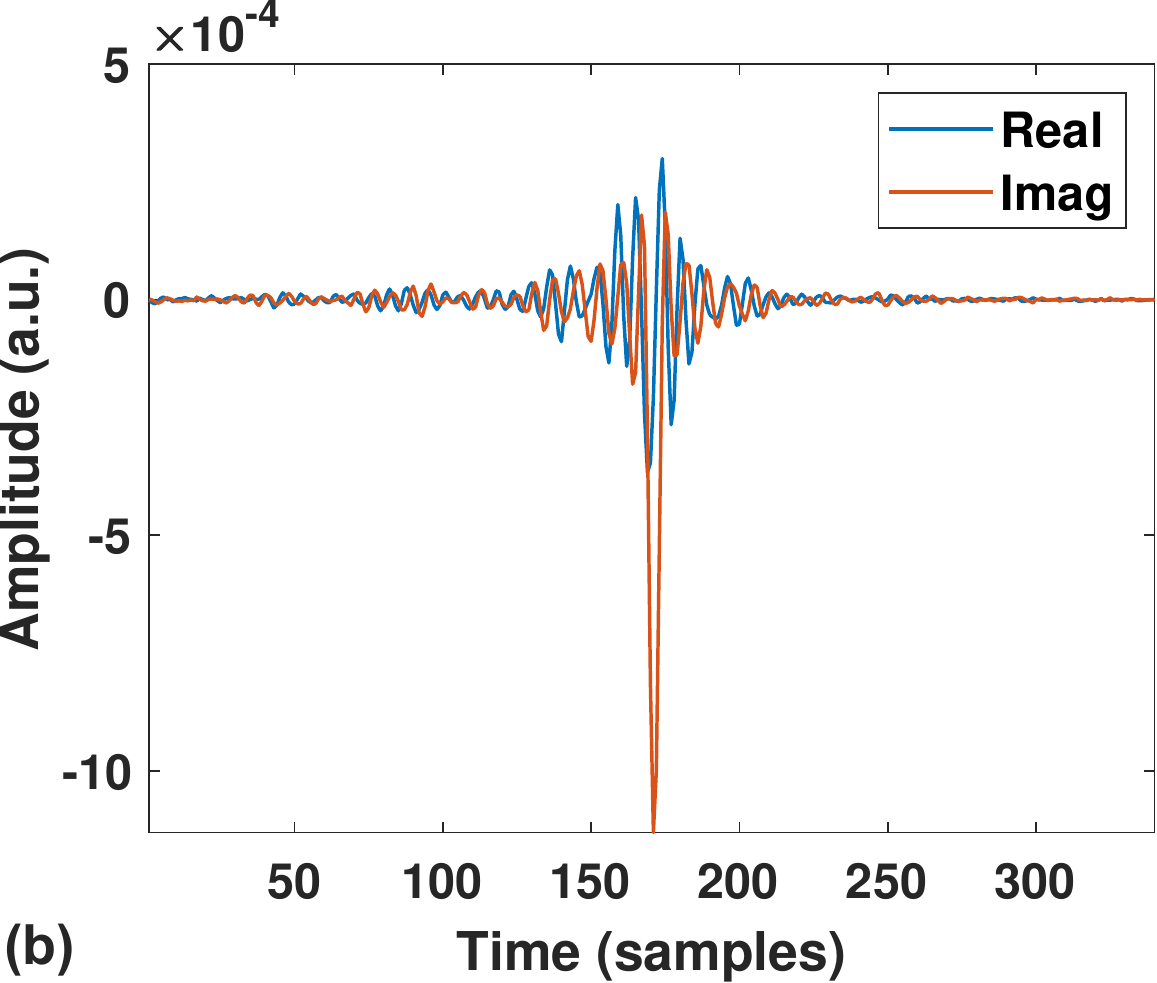}
    \includegraphics[width=0.34\textwidth]{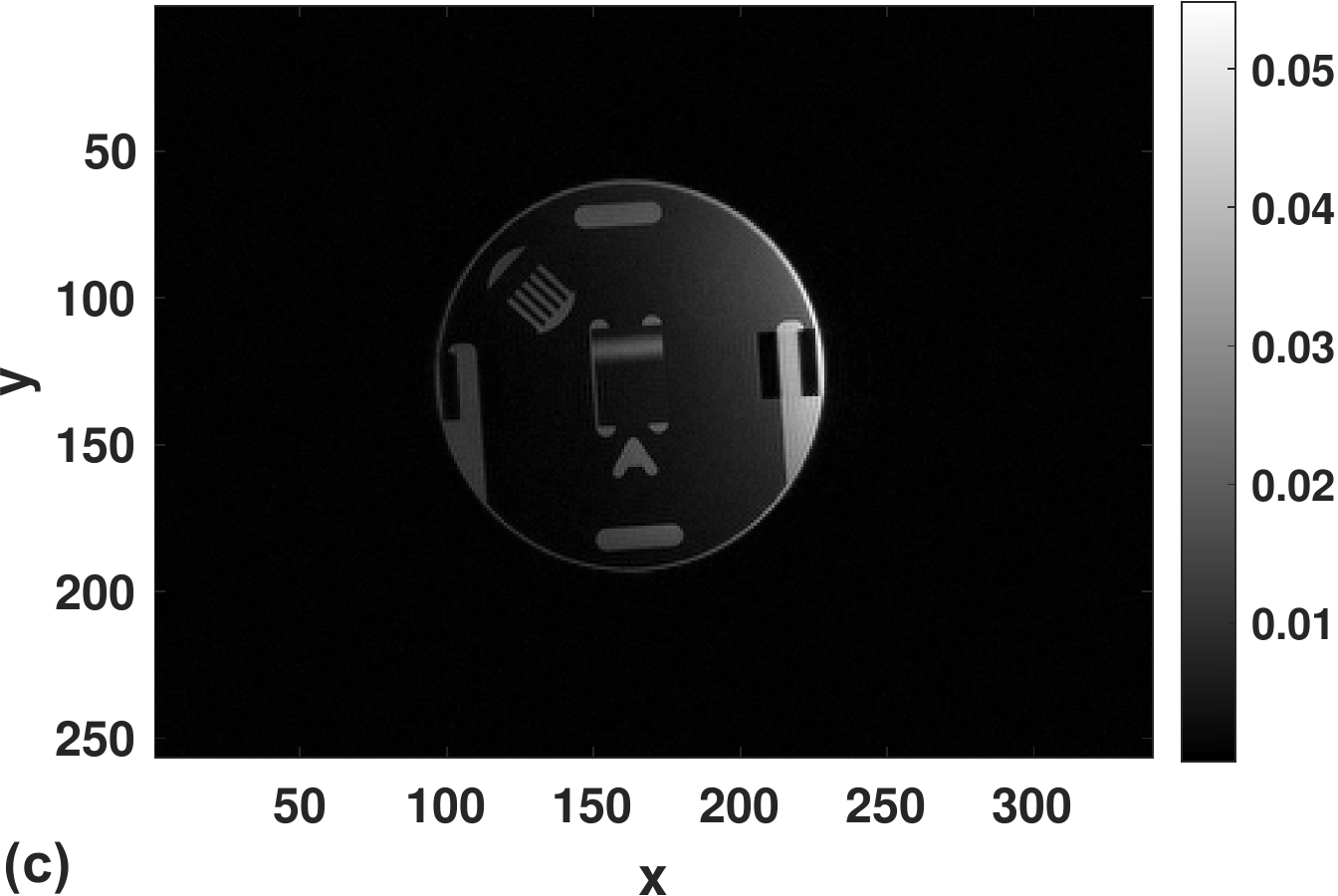}
    \caption{Typical 3T MRI data from an imaging phantom, acquired using a turbo spin echo (TSE) sequence and phase-frequency ($y$-$x$) encoding: (a) $k$-space magnitude, with the arrow indicating the center ($k_{x0},k_{y0}$); (b) complex time-domain signal corresponding to the horizontal line through the center (i.e., $k_{y}=k_{y0}$); and (c) reconstructed image using a 2-D fast Fourier transform. Raw data was kindly provided by~\cite{haldar}.}
    \label{fig:k_space_img}
\end{figure*}

In fact, despite the drawbacks of electromagnetic coupling, coaxial cables do have the advantage of maintaining sufficient DR for transmitting MRI signals. In extreme cases, the required DR can exceed 100~dB due to high signal amplitudes near the center of $k$-space~\cite{gabr2009mri,behin2005dynamic}. However, DR values $<88$~dB are sufficient for typical human scans at 3~T with multi-coil head arrays and birdcage head coils, while 75~dB DR is sufficient for cardiac and spine coil arrays~\cite{behin2005dynamic}. As an example, Fig.~\ref{fig:k_space_img}(a) shows typical $k$-space magnitude data from a 3~T scanner; the DR is $\sim$90~dB. Note that the data are acquired using conventional phase-frequency ($y$-$x$) encoding, such that each horizontal line corresponds to time-domain signals from a single scan. Fig.~\ref{fig:k_space_img}(b) shows the complex time-domain signal for the horizontal line passing through the center of $k$-space; this particular scan requires the largest receiver DR because of the large amplitudes present near the middle of the acquisition window. For completeness, Fig.~\ref{fig:k_space_img}(c) also shows the reconstructed image corresponding to the data in Fig.~\ref{fig:k_space_img}(a).

Existing wireless and optical RF links only provide 60-70~dB DR for full bandwidth MRI signals due to electrical and optical nonlinearities~\cite{aggarwal2016millimeter, riffe2014wireless, riffe2013identification, heid2009cutting, wei2007realization, scott2005wireless, yuan20084, memis2008miniaturized, biber2008analog, yuan2007direct, koste2005optical, du2007comparison, tang2015home, possanzini2011scalability}; based on the discussion above, this is clearly insufficient for regular MRI scans unless additional processing steps (e.g., signal compression or automatic gain control)~\cite{bollenbeck2005high,elliott1998improved} are used to reduce the DR. Fortunately, digital RF-over-fiber links can adopt on/off modulation to avoid degradation of DR due to nonlinearities~\cite{yang2010multichannel,gamage2009design}. Thus, they are attractive for transmitting RF signals in MRI~\cite{possanzini2011scalability,tang2015home}.

Delta-sigma modulation (DSM) converts an analog signal into a high-frequency low-resolution digital signal that can be easily modulated on optical fiber. DSMs are gaining popularity in data converter designs due to their cost-efficiency and lower circuit complexity compared to conventional analog-to-digital converters (ADCs)~\cite{gao2019digital, wang2016delta}. A delta-sigma ADC consists of the modulator and a digital low-pass filter (decimator) to generate the final low-data-rate high-resolution output; such ADCs are particularly useful in applications with low signal bandwidth and high signal resolution. The MR signal from a typical 3~T scanner has a bandwidth of $\sim$1~MHz (centered around a Larmor frequency of 123-128~MHz) and a typical DR between 75~dB and 88~dB, both of which make DSMs a very promising digitization method for MRI scanners. These advantages have also led to recent work on DSM-based digital RF-over-fiber links for cellular networks~\cite{jang2017digital}. Additionally, DSM-based fiber links also allow very simple digital-to-analog conversion (DAC) at the receiver; the analog signal can be recovered simply by an analog or digital low-pass filter because of the pulse-density modulated nature of DSM outputs, whereas conventional ADCs require a high-resolution DAC for signal reconstruction. Note that the analog signal must be recovered before interfacing the proposed RF-over-fiber link with commercial MRI consoles, which expect analog inputs.

When everything is taken into consideration, the signal integrity and the DR of the transmission link must have first priority in order to make it practically useful. Therefore, we propose and prototype a new digital optical transmission system based on delta-sigma modulation that aims to provide high enough DR for regular MRI scans. The rest of the paper is organized around the necessary hardware and instrumentation, as follows. Section~\ref{sec:dsm} reviews the basic design considerations for DSM-based RF-over-fiber links. A low-bandwidth DSM-based digital RF-over-fiber link using discrete components is described in Section~\ref{sec:discrete}, while a high-bandwidth link using a custom DSM chip is described in Section~\ref{sec:custom}. Experimental results from both links are presented in Section~\ref{sec:expt}, while Section~\ref{sec:conclusion} discusses these results and concludes the paper.

\section{Design Considerations for RF-over-Fiber Links}
\label{sec:dsm}
This section reviews the key performance metrics of DSMs and then describes how they affect the overall DR of the link.

\subsection{Performance of Delta-Sigma Modulators}
\begin{figure}[htbp]
    \centering
    \includegraphics[width=\columnwidth]{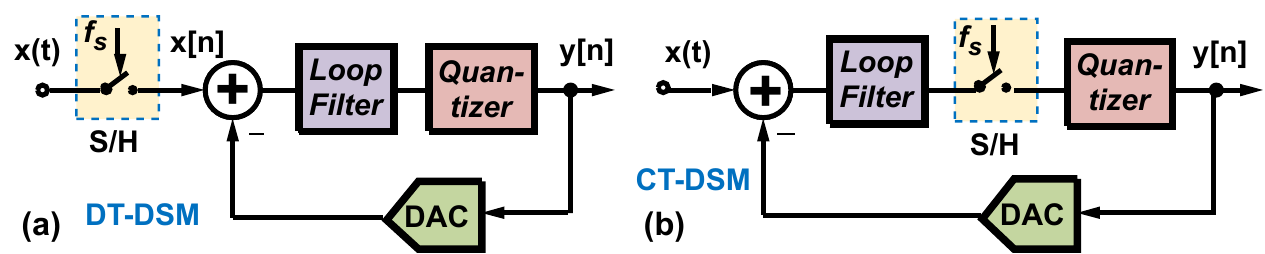}
    \caption{Basic DSM architectures: (a) discrete-time (DT-DSM), and (b) continuous-time (CT-DSM).}
    \label{fig:dsm_structures}
\end{figure}

Delta-sigma modulation relies on two key concepts to reduce the in-band quantization noise of a low-resolution quantizer: oversampling and noise shaping. Oversampling means that the input signal is sampled much faster than that required by the Nyquist criterion, while noise shaping alters the spectrum of the quantization noise to reduce its power in the signal band. For this purpose, the input signal is filtered by a loop filter, whose output is quantized and then fed back to the input through a DAC as shown in Fig.~\ref{fig:dsm_structures}. Conventionally, the sample-and-hold (S/H) is placed before the loop filter, resulting in a discrete-time DSM (DT-DSM) with a discrete-time loop filter (Fig.~\ref{fig:dsm_structures}(a)). Alternatively, the S/H can be placed after the loop filter, resulting in a continuous-time DSM (CT-DSM) with a continuous-time loop filter (Fig.~\ref{fig:dsm_structures}(b)).

Both types of DSMs can be designed for either low-pass or band-pass input signals by using integrators or resonators in the loop filter, respectively~\cite{pavan2017understanding}. MRI signals are narrowband and centered around the Larmor frequency $f_{L}$, i.e., are band-pass in nature with $f_{L}\gg f_{B}$, the signal bandwidth; thus, band-pass DSMs (BP-DSMs) seem to be a natural choice. However, the pass-band of a typical BP-DSM is centered around $f_{s}/4$ where $f_{s}$ is the sampling frequency. Thus, using a BP-DSM imposes the condition $f_{s}=4f_{L}$, which results in very high sampling frequencies for such narrowband signals. For example, we would need $f_{s}\approx 500$~MHz for signals from a 3~T scanner, resulting in a challenging circuit design with high power consumption. Hence, we focus here on down-converting the MRI signal and using a low-pass DSM (LP-DSM) for which $f_{s}$ (and thus OSR) can be set independent of $f_{L}$.

Given a low-pass input signal (obtained after down-conversion) that is confined between DC and the bandwidth \(f_B\), the oversampling rate (OSR) is defined as the ratio between \(f_s\) and the Nyquist frequency 2\(f_B\):
\begin{equation}
OSR = \dfrac{f_s}{2f_B}.
\label{eq:1} 
\end{equation}
Theoretically, the variance of in-band quantization noise $\sigma^{2}_{Q}$ is determined by both OSR and the noise transfer function (NTF) of the modulator, as follows:
\begin{equation}
    \sigma^{2}_{Q}=\frac{\Delta^{2}}{\pi}\int_{0}^{\pi/OSR}{\left|NTF(\omega)\right|^{2}d\omega}.
\label{eq:sigma_q}
\end{equation}
Here $\Delta$ is the step size of the quantizer, i.e., the least significant bit (LSB). An $N^{\mathrm{th}}$-order LP-DSM uses $N$ integrators within the loop filter. The resulting NTF is a high-pass function $\propto \omega^{N}$ within the signal band~\cite{aziz1996overview}, such that:
\begin{equation}
\sigma^{2}_{Q}\propto OSR^{-(2N+1)}.
\label{eq:2}
\end{equation}
Consequently, increasing the sampling rate can significantly increase the signal-to-quantization noise ratio (SQNR) $P_{sig}/\sigma^{2}_{Q}$, where $P_{sig}$ is the signal power. In particular, $SQNR$ (in dB) is proportional to $(2N+1)~\cdot 10\log_{10}{OSR}$. For example, doubling OSR in a second-order DSM will result in an SQNR increase of approximately 15~dB. However, the overall output SNR and resolution of the modulator is also affected by other sources of noise, notably i) thermal and $1/f$ noise from the circuits, and ii) clock jitter. Expressing the variance of these error sources as $\sigma^{2}_{N}$ and $\sigma^{2}_{J}$, respectively, the SNR is:
\begin{equation}
SNR=\frac{P_{sig}}{\sigma^{2}_{Q}+\sigma^{2}_{N}+\sigma^{2}_{J}}.
\label{eq:snr}
\end{equation}
Here we have assumed that the three error sources are uncorrelated, which is generally a safe assumption. The maximum output SNR for a DSM is typically limited by the onset of instability at large input amplitudes.

In addition to random error sources, DSMs generate signal-dependent distortion components (harmonics and intermodulation products) at large input amplitudes. The signal-to-noise and distortion ratio (SNDR) takes into account the degradation in SNR due to distortion, and is given by:
\begin{equation} 
SNDR = \dfrac{P_{sig}}{\sigma^{2}_{Q}+\sigma^{2}_{N}+\sigma^{2}_{J}+P_{D}}.
\label{eq:4}
\end{equation}
Here $P_{D}$ is the power of all distortion components. Computationally, the noise and distortion is the sum of all non-fundamental signals up to half of the sampling frequency (i.e., $f_s/2$), including harmonics but excluding DC. As the input signal amplitude $A$ increases, $P_{sig}$ increases as $A^{2}$, while $P_{D}$ increases at a much faster rate (e.g., as $A^{4}$ for the second harmonic). Thus, SNDR reaches its peak value SNDR$_{max}$ for a particular input signal amplitude $A_{max}$ and then decreases. 

Given that MRI receivers acquire wideband time-domain signals (as shown in Fig.~\ref{fig:k_space_img}(b)), using frequency-domain metrics such as SNDR results in overly-pessimistic estimates for receiver DR. Note that the frequency domain corresponds to the image dimension, which typically has much lower SNR$_{max}$ than the time-domain. Thus, small distortion components can be ignored since they are invisible in the image. Assuming the receiver is non-adaptive, the maximum output SNR (denoted by SNR$_{max}$) is thus a more appropriate measure of DR, and the DSM must be designed to ensure SNR$_{max}\geq$ DR, the target dynamic range of 80-90~dB. However, SNR$_{max}$ requirements can be significantly reduced by using adaptive receivers based on either signal compression~\cite{bollenbeck2005high} or automatic gain control~\cite{elliott1998improved}.

\subsection{Estimation of Linearity and Dynamic Range}
Regardless of the transmission method, the output RF signal amplitude at the receiver should be linearly related to the input RF signal amplitude to prevent image distortion. Deviations from input-output linearity occur both when the RF input power is too high or too low. The linearity at high RF power is usually limited by the linear range of the DSM, which determines SNR$_{max}$. The linearity at low RF power is usually limited by quantization noise components that are not truly random but correlated with the input. Such correlated components give rise to spurious ``idle tones'' in the output spectrum, some of which lie within the signal band~\cite{pavan2017understanding}. These can be reduced by using a multi-bit quantizer, adding dithering~\cite{reiss2008understanding}, or using a higher-order modulator.

End-to-end tests of the link (i.e., with RF inputs and RF outputs) allow the useful DR to be determined as the linear region where the signal is i) minimally distorted during transmission, and ii) has SNR $>1$. In particular, the input-output response for large signals can be assessed using various metrics to determine the linear range of the link; here we use the 1~dB gain compression point ($P_{1dB}$), which is defined as the signal power at which the output falls 1~dB below the linear response. Thus, the \emph{end-to-end DR} is defined as the range between where the RF signal hits the noise floor (set by both true noise sources and idle tones) and the input-referred $P_{1dB}$ point (known as $P_{1dB,in}$). Note that if the distortion is primarily generated by a memory-less third-order nonlinearity (often a good approximation), 1~dB gain compression corresponds to a SNDR of $\approx 16$~dB, which is barely acceptable for typical images. If higher-SNR images are anticipated, the linear range can be re-defined using a smaller amount of gain compression (e.g., 0.1~dB), resulting in a small decrease in DR.

\section{Prototype Link Using Off-the-Shelf Parts}
\label{sec:discrete}
This section describes the first version of our proposed digital RF-over-fiber link, which was designed and prototyped using commercial off-the-shelf (COTS) components.

\subsection{System Architecture}
\begin{figure}[htbp]
    \centering
    \includegraphics[width=1\columnwidth]{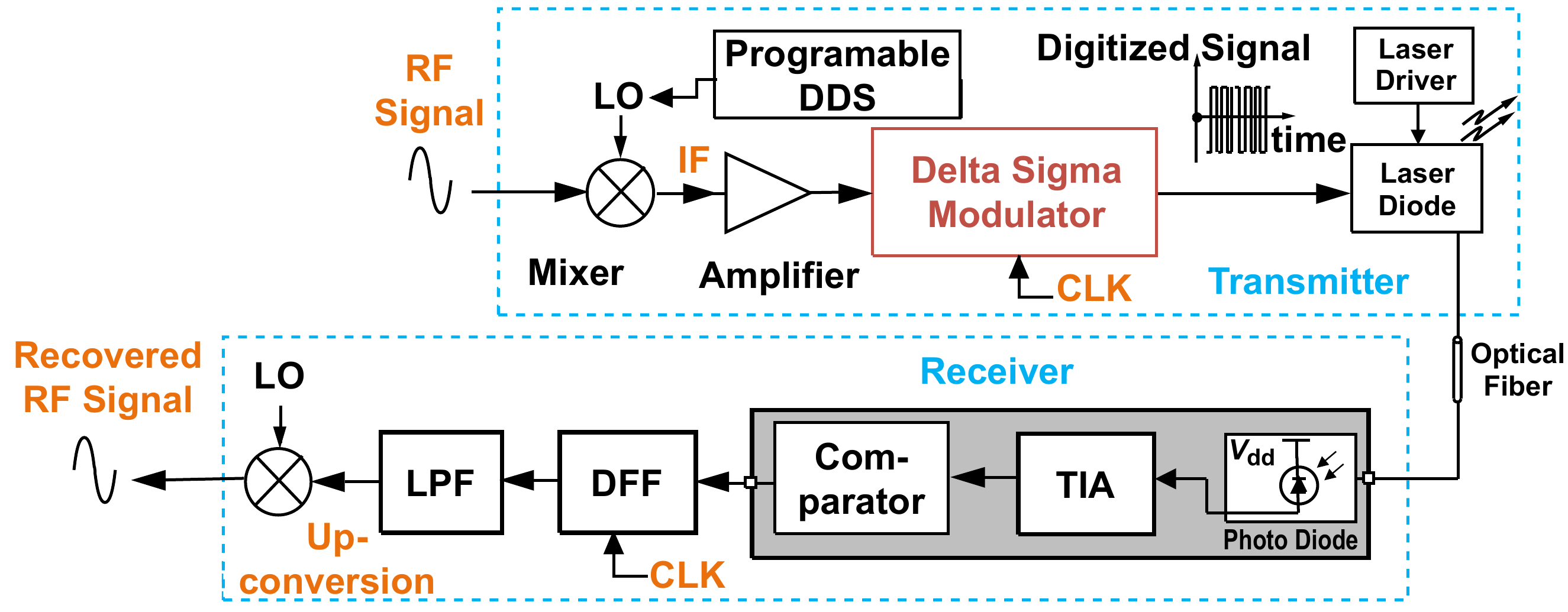}
    \caption{Block diagram of the basic digital RF-over-fiber link.} 
    \label{fig:1}
\end{figure}

The block diagram of the DSM-based digital optical link is shown in Fig.~\ref{fig:1}. The transmitter and receiver are connected through a single-mode optical fiber. The key components in the transmitter include a mixer for frequency down-conversion, a direct digital synthesizer (DDS) to generate the local oscillator (LO) for the down-conversion, an LP-DSM with 1-bit quantizer for digitizing the down-converted signal, and a laser diode (LD) driver for on-off keying (OOK) modulation of light intensity on the optical fiber. The receiver contains a PIN photodiode (PD) and transimpedance amplifier (TIA) for detecting the transmitted light, a comparator to digitize the detector output, a D-type flip flop (DFF) to re-time the comparator output, and an analog low-pass filter and up-converting mixer to regenerate a band-pass MRI signal that can be directly fed into commercial MRI consoles. Note that the final stages in the receiver (the analog filter and up-converter) can be eliminated if the MRI console has the option to directly store and process the digitized baseband signals\footnote{Major MRI manufacturers generally do not provide this option, but it is easy to realize within custom designs.}.

The commercial DSM used in this system is the AD7403 (Analog Devices) DT-DSM. Since this is a second-order low-pass design, the analog RF input is first down-converted to the baseband using a wideband frequency mixer (ASK-2-KK81, Mini-Circuits). The LO signal used for frequency mixing is provided by a DDS (AD9910, Analog Devices), which is controlled by an on-board micro-controller (Teensy 3.6) programmed using Arduino IDE. The frequency mixing results in a baseband signal between DC and the bandwidth \(f_B\) along with other unwanted signals at higher frequencies. A low-pass filter is applied to filter out the signals outside the baseband. The resulting baseband signal is amplified and subsequently digitized by the DSM with a 20~MHz sampling clock, which is the highest sampling rate allowed by the AD7403. The DDS uses the same clock (in conjunction with an internal PLL-based clock multiplier) while generating the LO. 

In our implementation, the receiver maintains synchronization with the transmitter by broadcasting the sampling clock via a separate fiber-optic or coaxial link (not shown); note that a single clock link is sufficient for a large RF coil array, such that its overhead is relatively small. However, while this method maintains frequency synchronization, there can be non-negligible timing error between the transmit and receive clock timing due to mismatched propagation delay (5-5.5~ns per meter of fiber) between the data and clock links. As a result, the DFF in the receiver uses a time-delayed version of the received clock to re-time the recovered data, with the delay adjusted via a digital delay line (not shown in the figure) to maximize the timing margin.

\begin{figure}[htbp]
    \centering
    \includegraphics[width=0.9\linewidth,height=3cm]{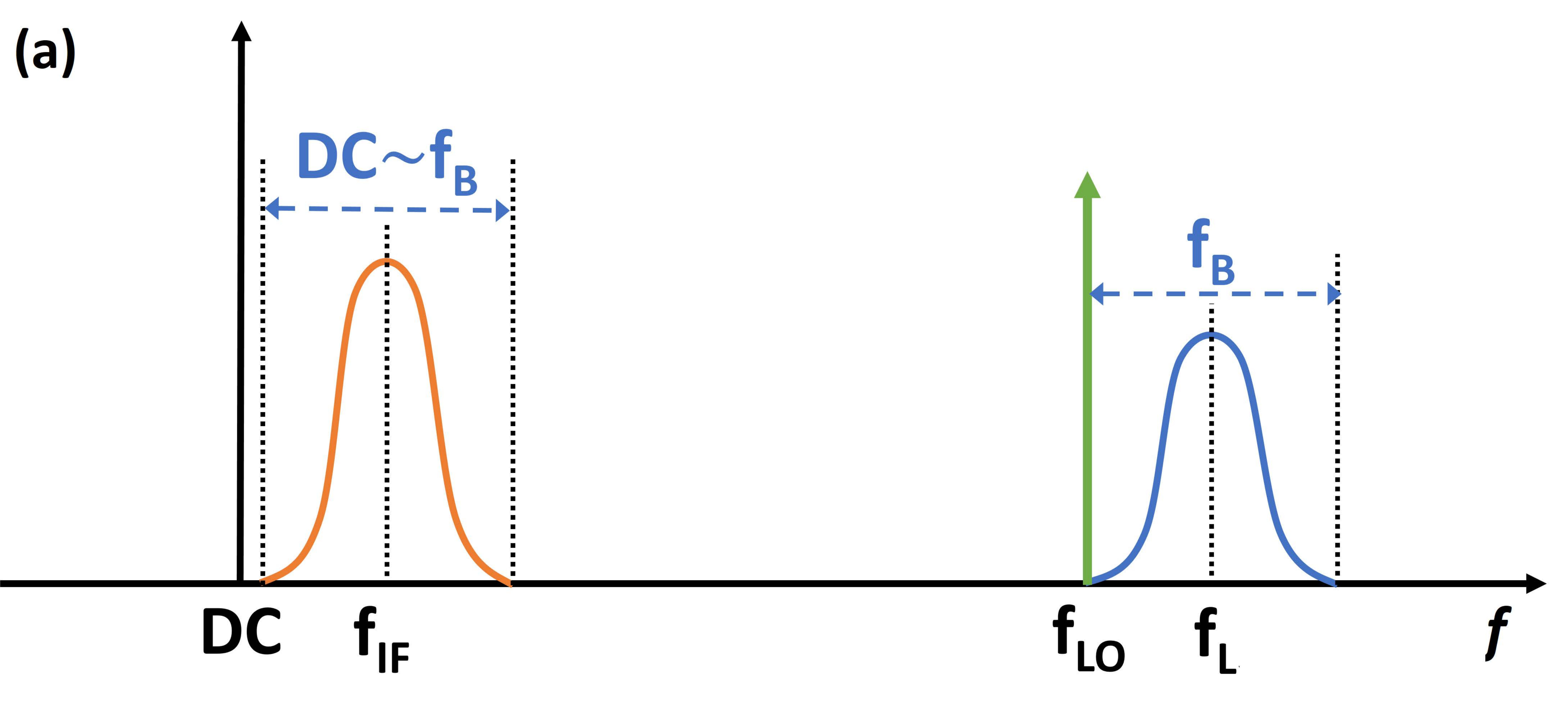}
    
    \includegraphics[width=0.9\linewidth, height=3cm]{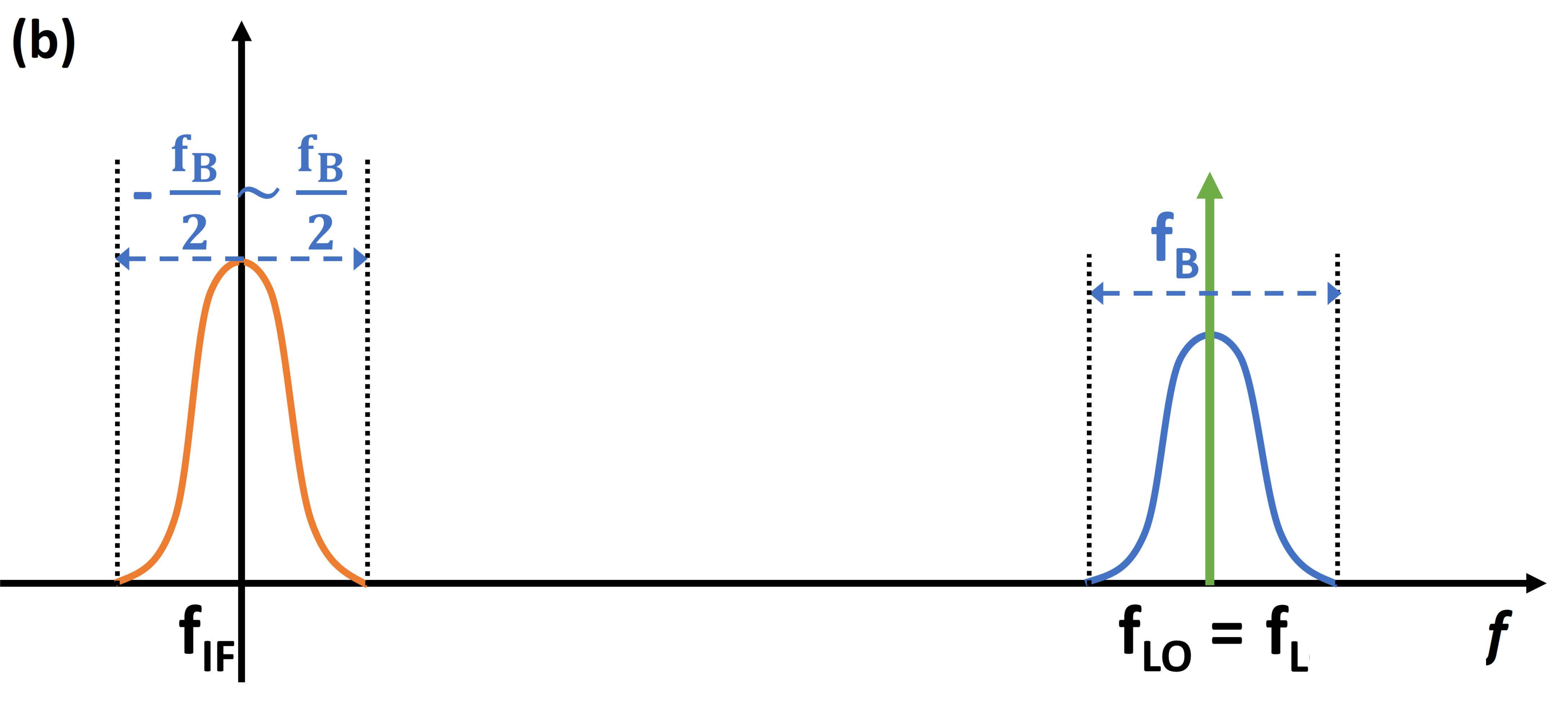}
    \caption{Spectra of MRI signals prior to digitization in two frequency down-conversion scenarios: (a) non-zero IF using a single mixer (only one sideband shown for simplicity), and (b) zero IF using a quadrature (I-Q) mixer. Here $f_{B}$, $f_{L}$, $f_{LO}$, and $f_{IF}$ refer to the signal bandwidth, Larmor frequency, local oscillator (LO) frequency, and intermediate frequency (IF) after down-conversion, respectively.} 
    \label{fig:2}
\end{figure}

The intermediate frequency (IF) of the MRI signal after down-conversion is $f_{IF}=\left|f_{L}-f_{LO}\right|$, where $f_{LO}$ is the LO frequency. Selection of the LO and IF frequencies presents another important design issue. Fig.~\ref{fig:2}(a) shows the case with a single mixer and $f_{IF}\approx f_{B}/2$, which ensures that the upper and lower sidebands do not overlap after down-conversion. In this case, OSR is defined by (\ref{eq:1}). However, OSR can potentially be doubled if a quadrature mixer is used, since in this case one of the sidebands is cancelled, which allows the spectrum of the down-converted signal to be centered at DC ($f_{IF}=0$, as shown in Fig.~\ref{fig:2}(b)). However, the resulting single-sideband transmitter would need two DSMs and then either i) two independent LD drivers and optical fibers, or ii) a wavelength-division multiplexing (WDM) network for transmitting both in-phase and quadrature (I and Q, respectively) signal components on the same fiber. Our prototype link uses a single mixer design to simplify the hardware implementation. 

\subsection{Circuit Design}
\begin{figure*}[htbp]
    \centering
    \includegraphics[width=0.32\textwidth]{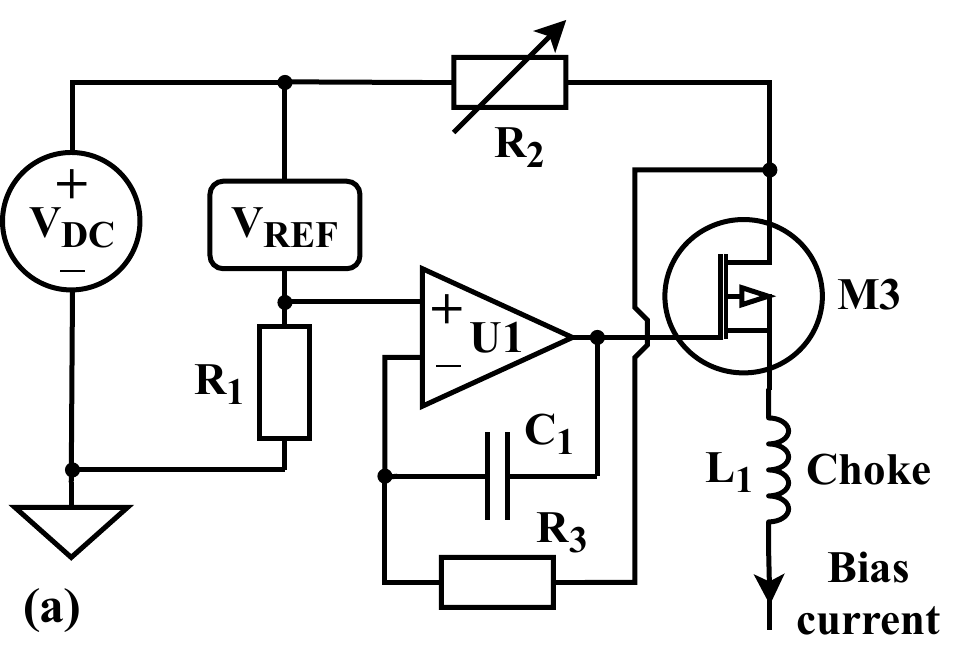}
    \includegraphics[width=0.27\textwidth]{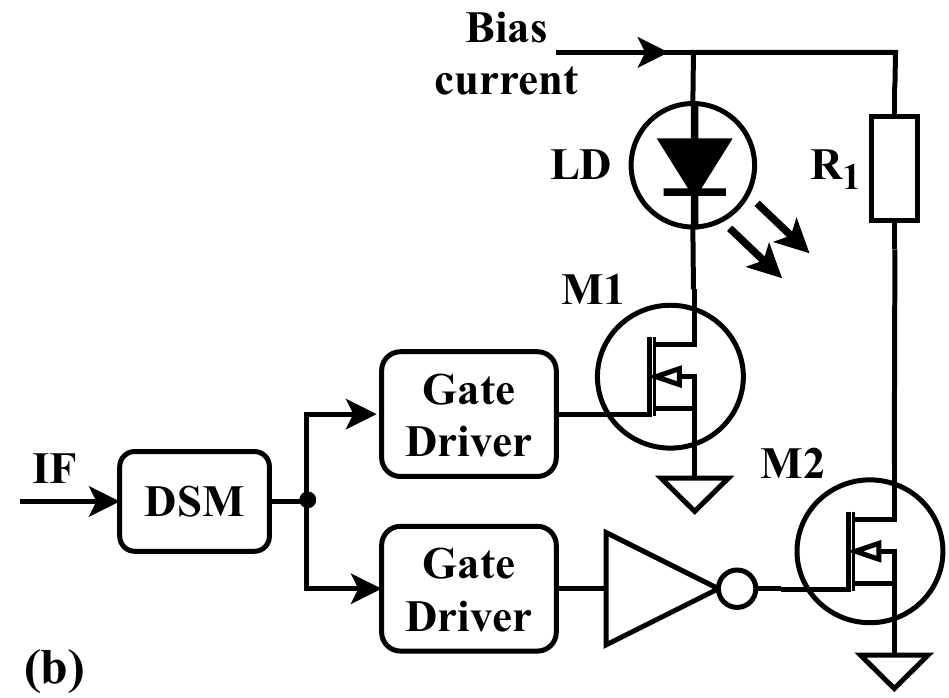}
    \includegraphics[width=0.32\textwidth]{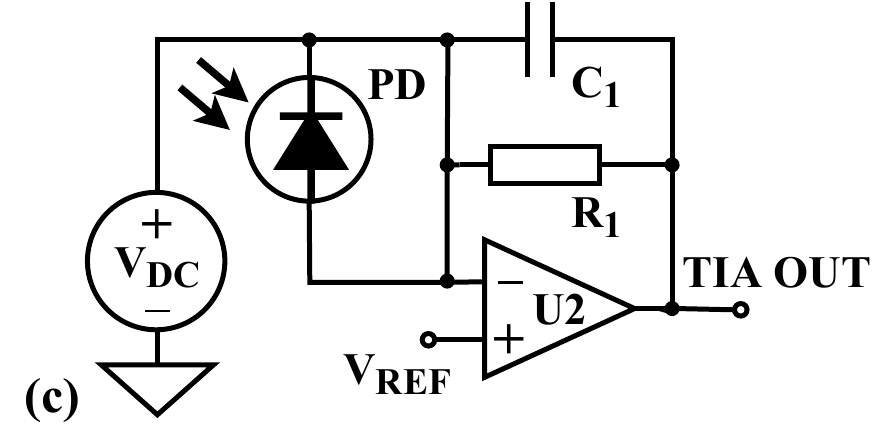}
    \caption{Schematics of the digital optical link: (a) bias current generator; (b) E/O converter for the DSM-digitized signal; and (c) O/E converter using a photodiode and a TIA.}
    \label{fig:3}
\end{figure*}

The transmitter contains a laser driver circuit (Fig.~\ref{fig:3}(a)) that provides the laser diode (FP-1310-5I-50SMF-FCUPC, Finisar) with its DC bias current. The circuit is designed with a 2.5~V voltage reference (\(V_{ref}\)), an operational amplifier (U1), a p-type MOSFET, and a potentiometer (denoted by \(R_2\)). Because the voltage across the potentiometer is determined only by the voltage reference, the bias current (\(I_{bias}\)) can be easily tuned by varying the potentiometer’s resistance, according to:
\begin{equation}
I_{bias} = \dfrac{V_{ref}}{R_2}.
\label{eq:3}
\end{equation}
The optical output is generated by a single-mode 1310~nm Fabry-Perot (FP) laser diode pigtailed with a single-mode optical fiber. The laser diode (LD) also contains a built-in monitoring photodiode that can be used to observe and/or regulate the optical output.

%
%
%

A schematic of the digital electrical-to-optical conversion (E/O) circuit is shown in Fig.~\ref{fig:3}(b). The cathode of the LD is only connected to the digitized baseband signal from the DSM, whereas its anode is connected to both the baseband signal and the DC current source. A choke is placed on the DC branch to prevent any high-frequency signals from flowing into the bias current generator. The digitized DSM output controls the two n-type MOSFET switches (M1 and M2). When the DSM outputs logic high, the MOSFET switch M1 in series with the LD turns on, so the bias current $I_{bias}$ flows through the LD and turns it on. When the DSM outputs logic low, M2 turns on, the bias current goes through the resistor $R_{1}$, and the LD is turned off. In summary, the digital output of the DSM controls the path of $I_{bias}$ to turn the LD on and off, which in turn generates an OOK-modulated optical signal. Note that this differential switching technique is preferred since it minimizes unwanted glitches due to charge injection.

The optical signal from the laser diode propagates through the optical fiber and is received by a pigtailed PIN photodiode (DPIN-23133, Precision Micro-Optics). The photodiode, which is reverse-biased to minimize capacitance, converts the light power back to current, which is subsequently converted to voltage by a TIA with a fixed transimpedance $R_{1}=1$~k$\Omega$ (Fig.~\ref{fig:3}(c)). The TIA was realized using a high-speed op-amp (LT1363, Analog Devices) with a gain-bandwidth product of 70~MHz. The recovered voltage signal is digitized by a comparator to minimize amplitude distortions introduced by LD power fluctuations and TIA noise, and then re-sampled on clock edges by a DFF to eliminate any accumulated timing jitter during the transmission. The clock used in the DFF is derived from the same source as that used in the DSM, as described in the previous section. The final digital-to-analog conversion is performed simply by an active analog low-pass filter (LPF). Finally, the recovered analog baseband signal is up-converted back to the Larmor frequency using a mixer. 

\section{High Bandwidth Link Using a Custom DSM}
\label{sec:custom}
The commercial DT-DSM chip used in Section~\ref{sec:discrete} (the AD7403) offers a maximum sampling frequency of 20~MHz. For MRI signals with \(\pm 500\)~kHz bandwidth, the highest OSR that can be provided by the AD7403 is 20. This results in an SNR$_{max}$ of approximately 52~dB, which is not sufficient for practical MRI applications. While a variety of other delta-sigma ADCs are available on the market, most of them have a built-in decimation filter to generate high-resolution outputs, and do not provide access to the low-resolution (e.g., 1-bit) output of the modulator required to drive the fiber. Thus, this section focuses on the design of a custom DSM chip that addresses the issues of limited sampling speed and SNR.

\subsection{Improved RF-over-Fiber System}
CT-DSMs have several advantages over DT-DSMs for realizing the relatively high-speed modulator required to obtain sufficient OSR. These include implicit anti-aliasing (due to the continuous-time low-pass loop filter), resistive input impedance (which simplifies driver requirements), and low-power operation~\cite{pavan2017understanding}. Thus, a custom high-frequency CT-DSM chip was designed for this application. The goal of the chip is to provide a signal bandwidth of 1~MHz using higher-order noise shaping than that available with a second-order DSM like the AD7403. Details of the circuit design were presented in our earlier work~\cite{gao2019digital}; here we provide a brief overview of the design and describe preliminary experimental results.

\begin{figure}[htbp]
    \centering
    \includegraphics[width=\columnwidth]{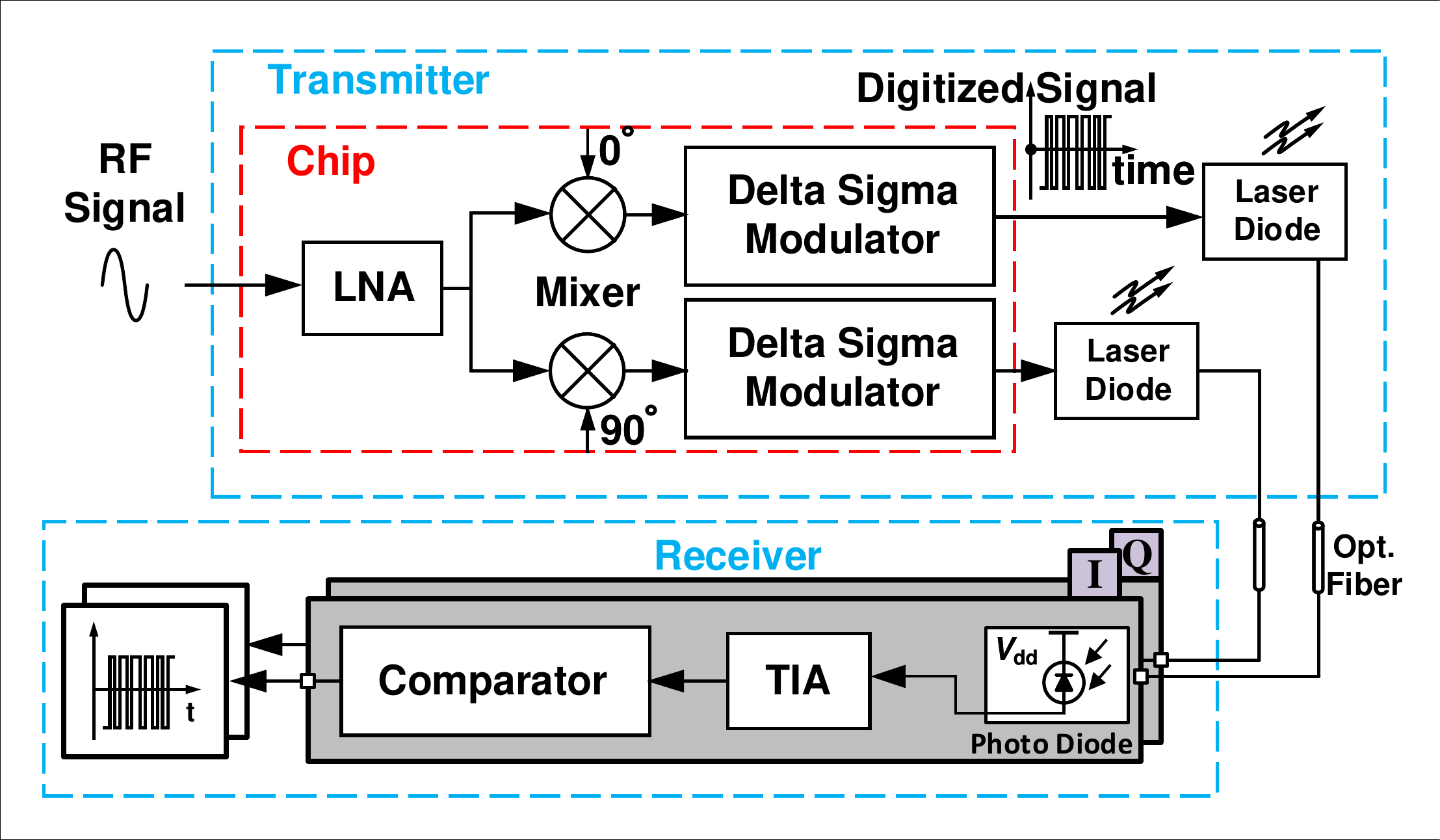}
    \caption{Block diagram of the high DR RF-over-fiber link.}
    \label{fig:improved_link}
\end{figure}

Fig.~\ref{fig:improved_link} shows a block diagram of the high DR version of the RF-over-fiber link. It differs from the prototype shown in Fig.~\ref{fig:1} as follows: i) the single mixer is replaced with a quadrature mixer to allow zero-IF down-conversion, thus doubling OSR for the same sampling rate; ii) the off-the-shelf DT-DSM is replaced by a custom CMOS chip containing two CT-DSMs (for digitizing the I and Q channels); and iii) two identical laser diode drivers, optical fibers, and receivers are used to independently process the I and Q signals.

\subsection{Parameters for the Custom DSM}
\begin{figure}
    \centering
    \includegraphics[width=0.75\columnwidth]{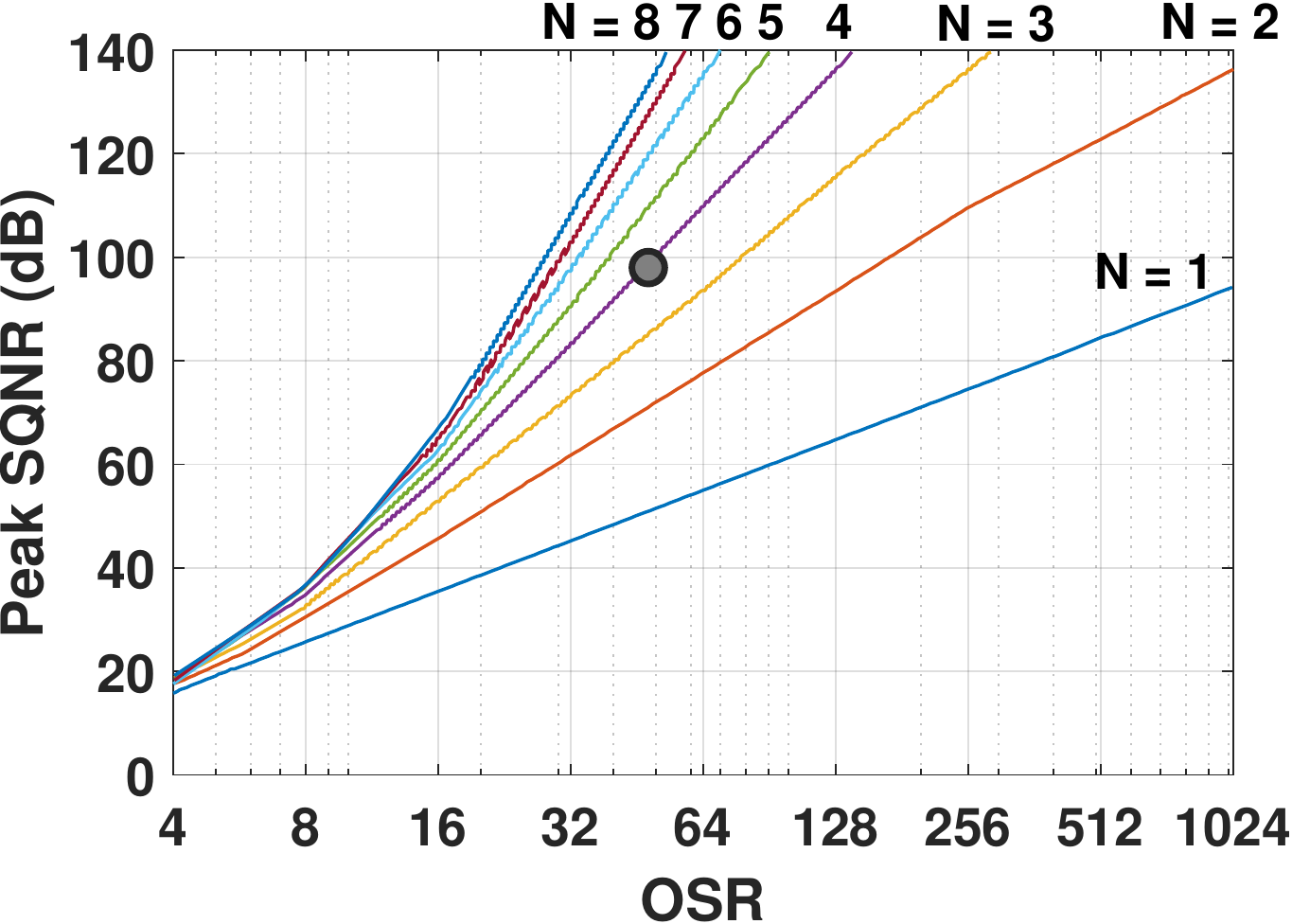}
    \caption{Available peak SQNR for DSMs of various orders ($N=1$ to 8) as a function of OSR. The chosen design ($N=4$, $OSR=50$) is represented by the gray dot. Figure generated using~\cite{schreier2021delta}.}
    \label{fig:sqnr_1bit}
\end{figure}

The main design parameters for a DSM include i) the number of quantization levels, $M$; ii) the number of integrators (i.e., the loop order $N$); and iii) the oversampling ratio (OSR). Here we limit ourselves to the $M=2$ case, i.e., binary or single-bit quantization. This is because the resulting two-level transfer functions for both the DAC and the optical link are inherently linear\footnote{Simply put, such linearity arises from the fact that two points define a straight line.}, which eliminates potential distortion mechanisms due to DAC mismatch and nonlinear E/O conversion. The choices of $N$ and OSR are based on a trade-off between circuit complexity and speed. For example, increasing $N$ results in more noise shaping, which allows OSR (and thus speed) to be reduced but at the cost of higher complexity. We explored these trade-offs using the widely-used open-source MATLAB Delta-Sigma Toolbox~\cite{schreier2021delta}, and the results are summarized in Fig.~\ref{fig:sqnr_1bit}. As shown in the figure, the peak SQNR available for a given OSR begins to saturate for $N>4$. Thus, we chose a fourth-order ($N=4$) CT-DSM for this application.

\begin{figure}[tbhp]
	\centering
	\includegraphics[width=0.85\columnwidth]{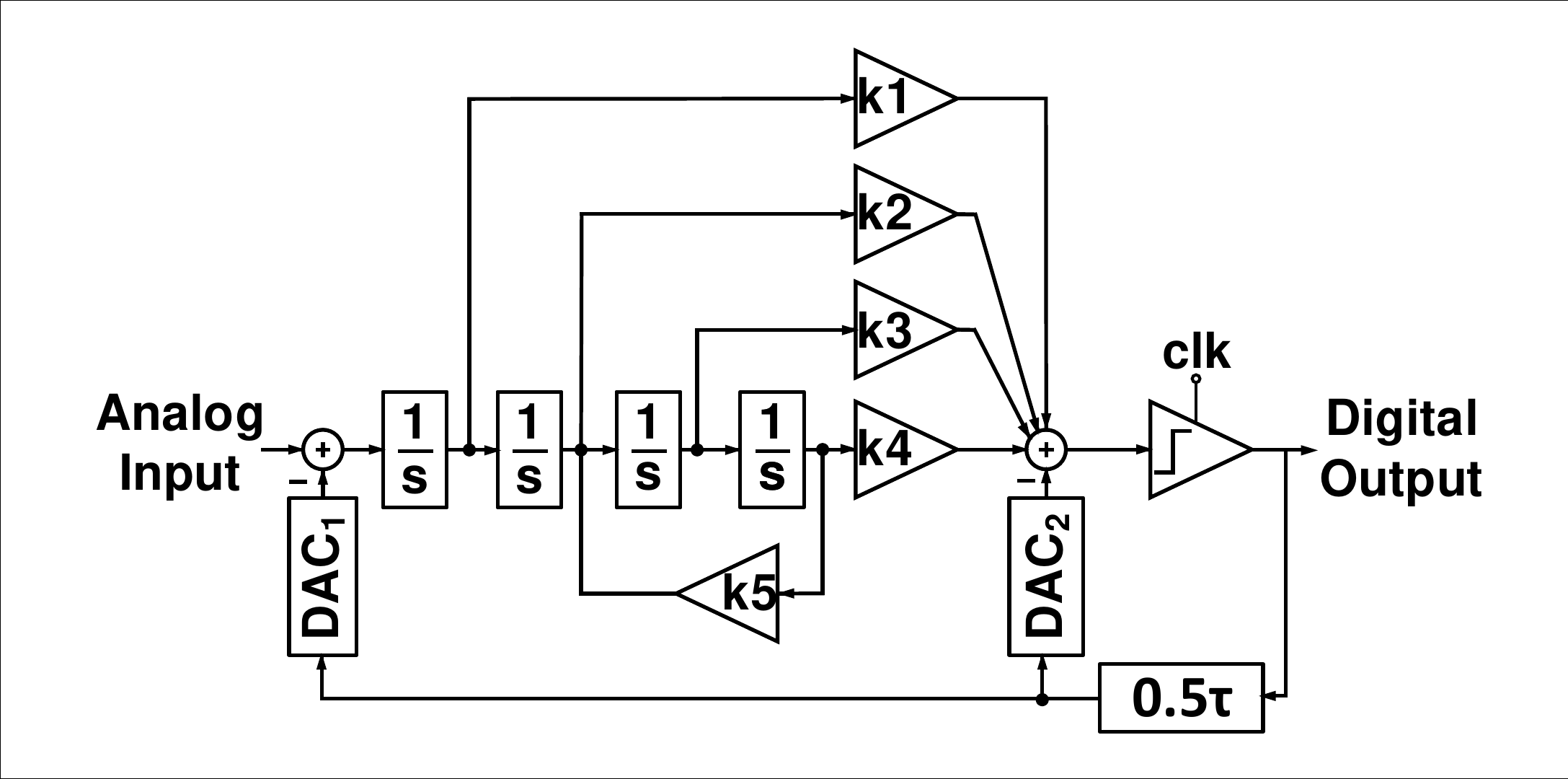}
	\caption{Block diagram of the proposed fourth-order CT-DSM with CIFF loop filter; $\tau=1/f_{s}$ is the clock period.}
	\label{fig:DSMblock}
\end{figure}

The loop filter incorporated the widely-used cascade of integrators feedforward (CIFF) topology, as shown in Fig.~\ref{fig:DSMblock}. The CIFF topology generates an NTF with passband zeros (which significantly improve peak SQNR for high-order modulators) by using distributed feedforward paths that drive a common summing node. We chose the CIFF topology for this design because it requires only two feedback DACs: $DAC_{1}$, which drives the input node; and $DAC_{2}$, which implements a feedback path around the quantizer that improves stability by partially compensating for the time delay of the loop filter. The design of the loop filter was further simplified by using a single feed-in path for the analog input. Also, its transfer function was optimized to maximize peak SQNR under the constraint $\|H\|_{\infty}\leq 1.5$ to ensure stability; here $\|H\|_{\infty}$ is the peak magnitude of the NTF versus frequency. The resulting closed-loop signal transfer function (STF) has a peak magnitude of 2.51 ($\approx 8.3$~dB), i.e., displays considerable ``peaking'' compared to its ideal value of 1. Such peaking is a generic feature of CIFF loop filters that use a single input feed-in path~\cite{pavan2017understanding}, and is generally undesirable because it results in higher signal swings at the integrator outputs (which in turn makes the circuit design of the loop filter more challenging). However, transistor-level simulations of the loop filter suggest that large enough signal swings can be generated without significant distortion, so the single feed-in design was retained.

An idealized MATLAB model of the chosen CIFF-based CT-DSM predicts SQNR$_{max}=95$~dB for $OSR=50$ (shown as the dot in Fig.~\ref{fig:sqnr_1bit}), which exceeds the target SNR; the corresponding sampling frequency is $f_{s}=100$~MHz. However, the modulator's SNR$_{max}$ is lowered by both thermal noise and clock jitter. In fact, most CT-DSMs are designed to ensure $\sigma^{2}_{N}\gg \sigma^{2}_{Q}$, which is beneficial for eliminating idle tones. Our design target was $\sigma^{2}_{N}\approx 4\sigma^{2}_{Q}$, such that $<20$\% of the total noise is due to quantization. This is a reasonable choice that results in SNR$_{max}=88$~dB in the absence of clock jitter (i.e., when $\sigma_{J}=0$). However, in practice clock jitter will further degrade SNR, as analyzed in the next sub-section.

\subsection{Effects of Clock Jitter}
Clock jitter presents one of the fundamental limitations to DSM performance, along with other sources of error such as integrator non-idealities (finite gain, slew rate, and nonlinearity)~\cite{boser1988design}. It is well-known that CT-DSMs are more sensitive to clock jitter than their discrete-time counterparts, with the dominant mechanism being random modulation of the feedback DAC pulse width~\cite{cherry1999clock,reddy2007fundamental}. Since such modulation occurs at the input terminal, random clock jitter adds a nearly frequency-independent (i.e., white) noise source to the DSM's output spectrum. Under fairly generic assumptions, the variance of the in-band noise due to this source is given by:
\begin{equation}
    \sigma_{J}^{2}=\left(\sigma_{T}f_{s}\right)^{2}\frac{\Delta^{2}}{\pi\cdot OSR}\int_{0}^{\pi}{\left|(1-e^{-j\omega})NTF(\omega)\right|^{2}d\omega}.
    \label{eq:sigma_j}
\end{equation}
Here $\sigma_{T}$ is the rms period jitter of the clock signal, which can be estimated from $L(f)$, the phase noise power spectrum of the clock, by using the well-known relationship~\cite{yamaguchi2001method}:
\begin{equation}
    \sigma_{T}^{2}=\frac{8}{\left(2\pi f_s\right)^{2}}\int_{0}^{f_s}{L(f)\sin^{2}\left(\pi f/f_s\right)df}.
\end{equation}
Here $f$ is the offset from the mean frequency $f_s$. Comparing eqns.~(\ref{eq:sigma_q}) and (\ref{eq:sigma_j}) reveals that while quantization noise is dominated by in-band behavior of the NTF, noise due to clock jitter is dominated by out-of-band behavior of the NTF.

In the proposed system, the off-chip clock source consists of a low-jitter crystal oscillator (AX3DCF1-200, Abracon) at a frequency of $2f_s=200$~MHz. The low-voltage differential signalling (LVDS) outputs of this oscillator drive a high-speed buffer/level translator chip (Si53306-B, Silicon Labs) that generates a CMOS-level waveform for driving the on-chip clock buffers. The estimated period jitter of the oscillator and buffer are 1.0~ps and 0.75~ps, respectively. Assuming these error sources are uncorrelated, the total expected clock jitter is approximately 1.2~ps. However, the actual jitter of the on-chip clock waveform may be significantly worse due to i) limited slew rates at the board/chip boundary due to parasitic capacitance from the package, and ii) additional jitter from the on-chip clock distribution network.

\begin{figure}[htbp]
    \centering
    \includegraphics[width=0.57\columnwidth]{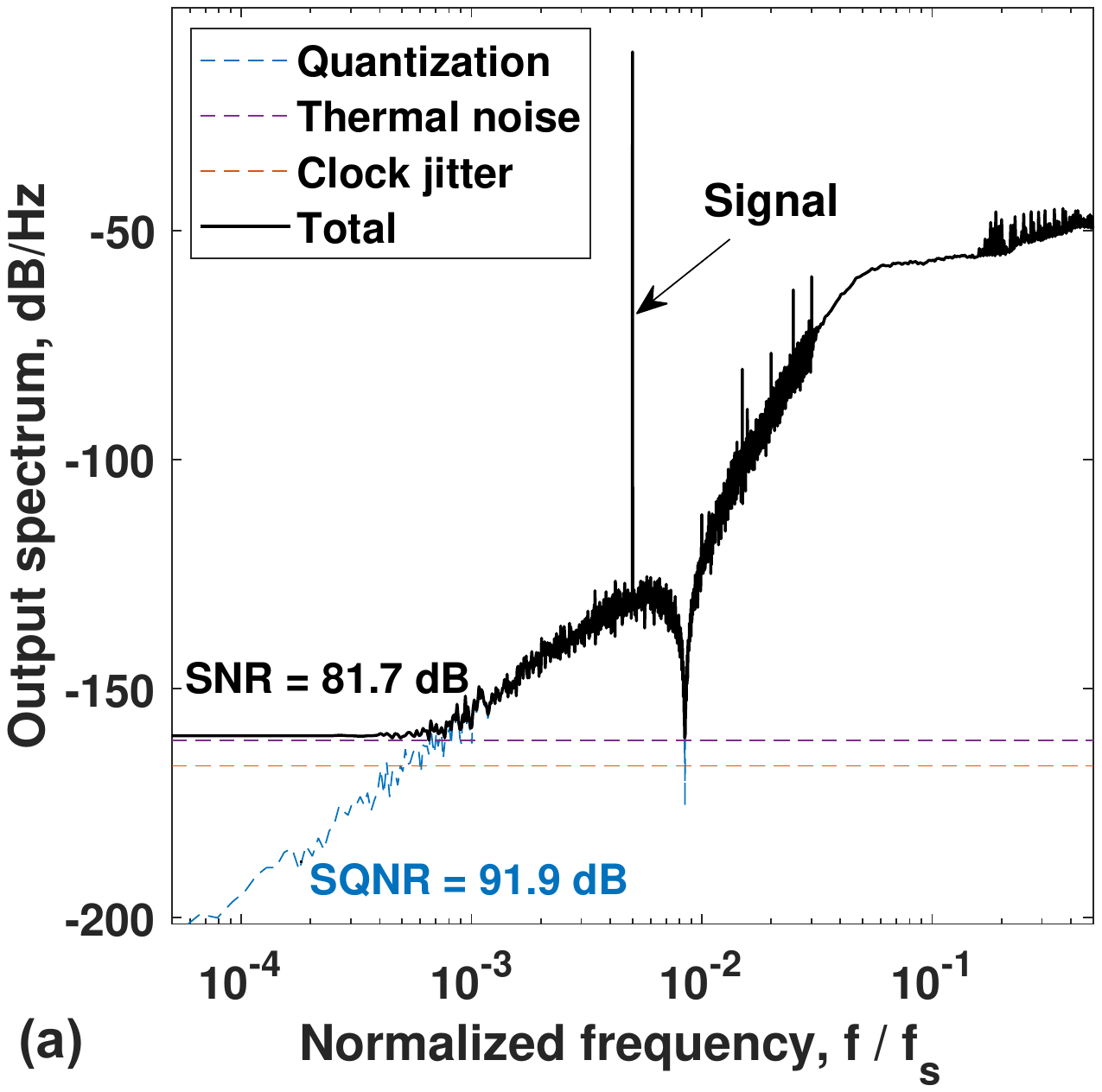}
    \includegraphics[width=0.41\columnwidth]{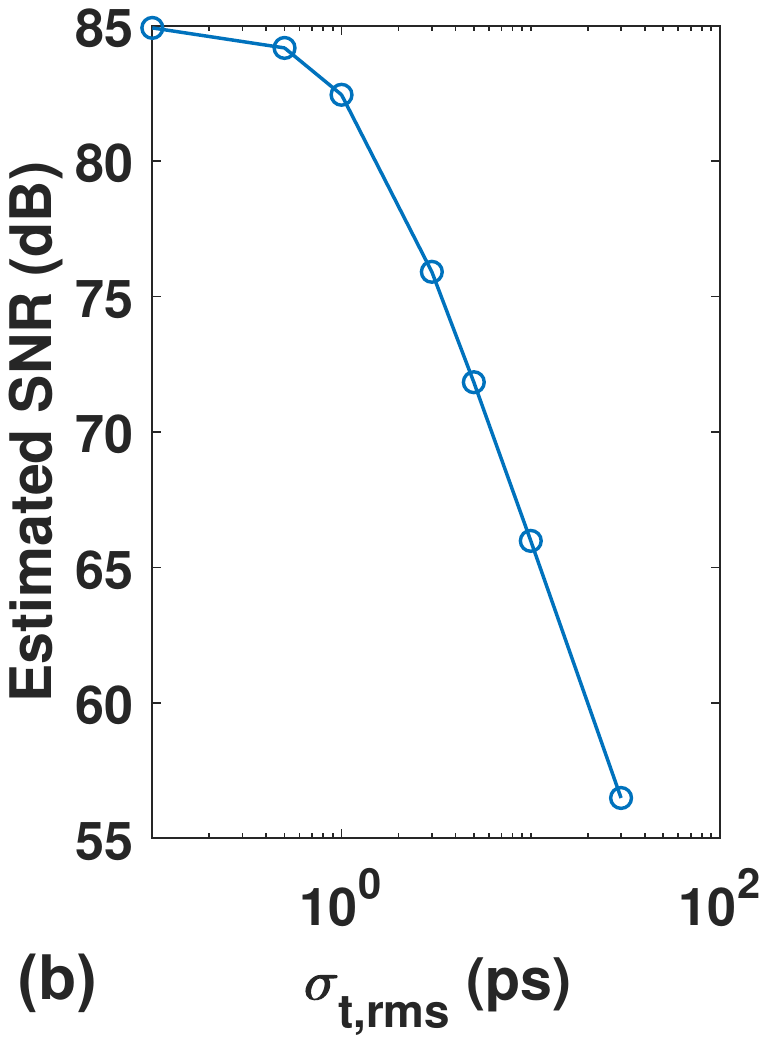}
    \caption{Simulated maximum SNR of the custom CT-DSM design: (a) output spectrum including both thermal noise and the nominal value of clock jitter ($\sigma_{T}=1.2$~ps); (b) SNR$_{max}$ as a function of $\sigma_{T}$ with other design parameters kept constant.}
    \label{fig:sim_snr}
\end{figure}

Fig.~\ref{fig:sim_snr}(a) shows the simulated output spectrum of the proposed CT-DSM design for a signal amplitude near $A_{max}$. The figure compares the spectra and SNR including i) only quantization noise, and ii) quantization noise, thermal noise, and clock jitter. Our results show that the expected amount of clock jitter (1.2~ps) degrades total noise by a modest $\sim$3.3~dB, resulting in SNR$_{max}=81.7$~dB. However, larger-than-expected clock jitter can cause much more significant degradation in SNR$_{max}$, as illustrated in Fig.~\ref{fig:sim_snr}(b).

\subsection{Circuit Design of the Custom CT-DSM} 
\begin{figure}
	\centering
	\includegraphics[width=0.95\columnwidth]{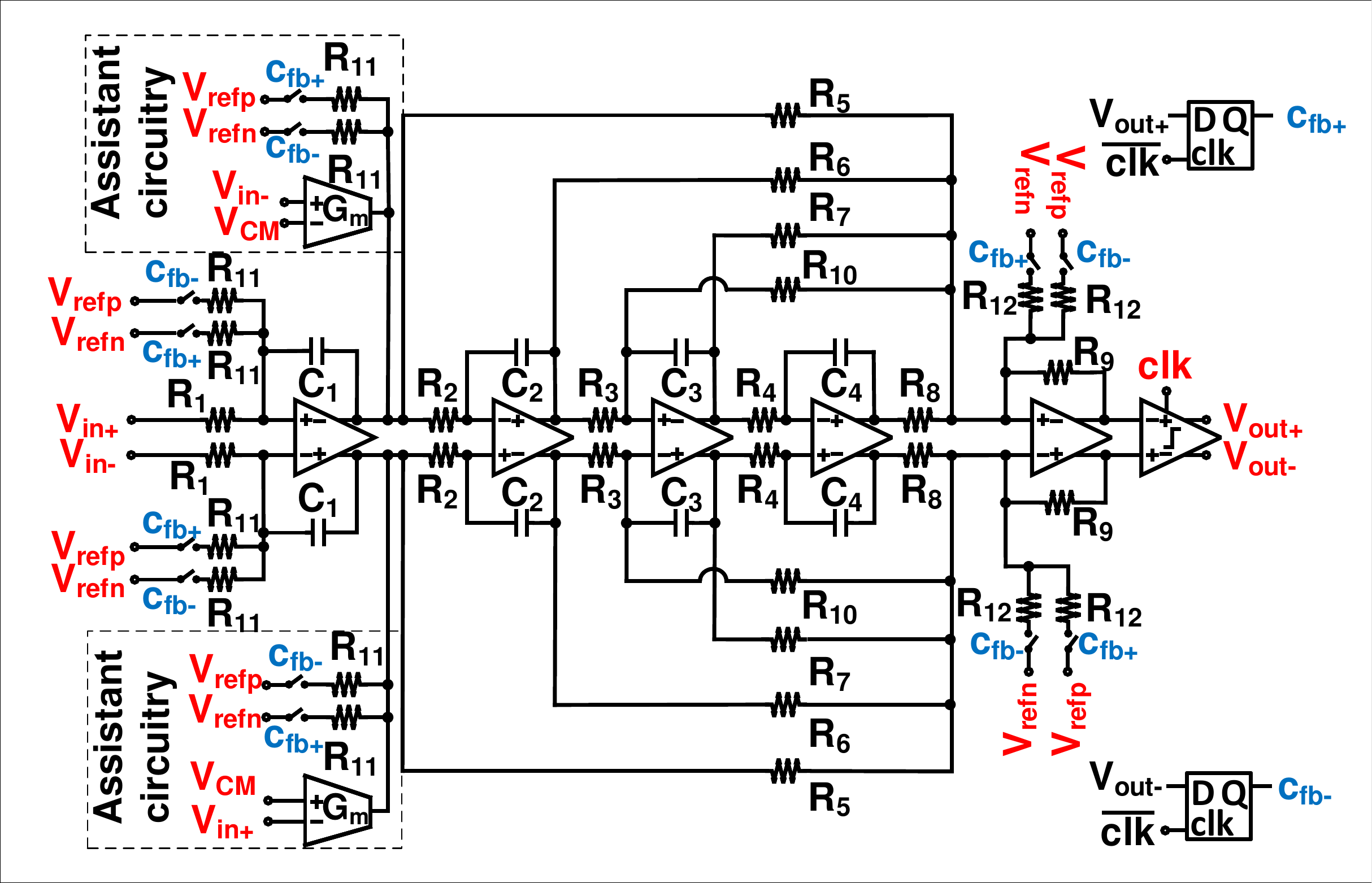}
	\caption{High-level schematic of the entire CT-DSM circuit.}
	\label{fig:DSM_sch}
\end{figure}

The custom DSM was designed in the TSMC 180~nm standard CMOS process. The high-level schematic of the whole circuit is shown in Fig.~\ref{fig:DSM_sch}. The loop filter uses an op-amp $RC$ topology. Four op-amps realize the four integrators required by a fourth-order loop, while a fifth op-amp sums up their outputs as required by the CIFF topology. The integration capacitors are realized as capacitive DACs to allow post-fabrication optimization of the NTF. The fully-differential op-amps use feedforward compensation, and their $1/f$ noise is eliminated via a chopping technique~\cite{billa2017analysis}. In addition, the linearity of the first integrator is enhanced via an additional feedforward ``assistant'' circuit (shown within the dashed lines in Fig.~\ref{fig:DSM_sch}) that reduces the op-amp's current source/sink requirements~\cite{pavan2010power}.

The output of the summing op-amp is digitized by a clocked dynamic comparator (acting as a 1-bit quantizer) and fed back via two one-bit non-return-to-zero (NRZ) resistive DACs, thus completing the DSM loop. The comparator uses the well-known StrongARM latch topology~\cite{razavi2015strongarm}.

The chip contains two matched CT-DSM circuits to process the down-converted I and Q channels. In addition, it contains i) a wideband class-AB follower to buffer the RF input signal, ii) double-balanced passive mixers for quadrature down-conversion, iii) a divide-by-2 circuit for generating the quadrature LO, iv) another divide-by-2 circuit to generate the chopping clock (which operates at $f_s/2$), and v) a standard 3-wire serial peripheral interface (SPI) port for programming the CT-DSM loop filter parameters.

\subsection{Simulation Results}
The chip was simulated at the transistor level using the Cadence Virtuoso suite. Each custom CT-DSM has a simulated maximum SNR (including thermal noise, but not clock jitter) of 89.3~dB for an input signal bandwidth of 1~MHz (\(OSR = 50\)), which is in good agreement with the theoretical model and should cover a broader spectrum of MRI applications. The simulated power consumption of each modulator is 8.8~mW, and the total power consumption of the chip (which is dominated by the two modulators) is approximately 18~mW.

\section{Experimental Results}
\label{sec:expt}
In this section we describe experimental results obtained from a complete version of the prototype RF-over-fiber link described in Section~\ref{sec:discrete}, as well as preliminary results obtained from the CT-DSM chip described in Section~\ref{sec:custom}.

\subsection{Prototype RF-over-Fiber Link}
\begin{figure}[htbp]
    \centering
    \includegraphics[width=1\columnwidth]{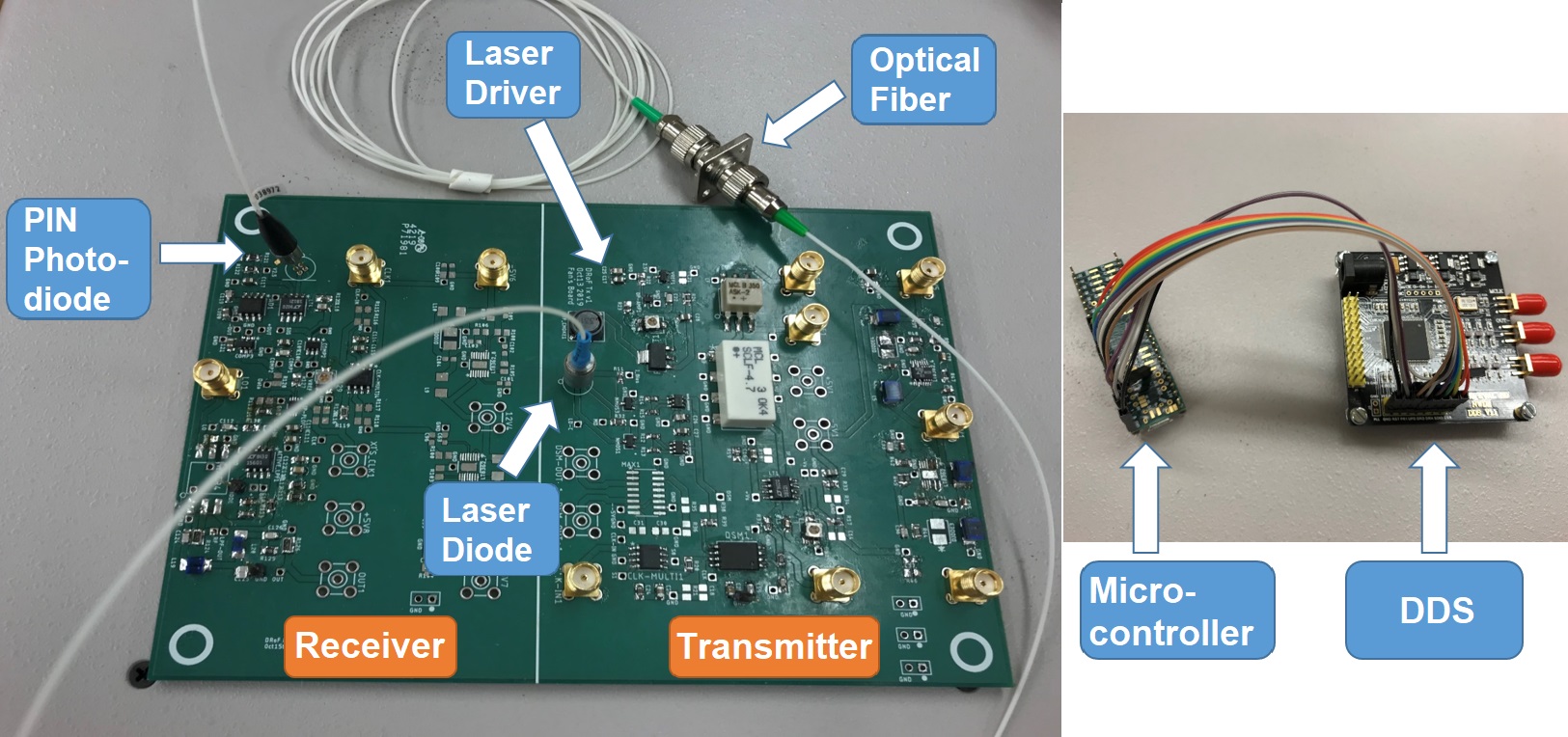}
    \caption{Implementation of the prototype RF-over-fiber link: (left) PCB containing the transmitter and receiver, and the optical fiber; (left) the DDS and associated micro-controller.} 
    \label{fig:4}
\end{figure}

Fig.~\ref{fig:4} shows a photograph of the proposed optical link, which was realized using i) a custom two-layer printed circuit board (PCB), and ii) a commercial DDS evaluation board (EVAL-AD9910, Analog Devices) connected to the Teensy 3.6 micro-controller. The latter is used to program the DDS via a standard four-wire SPI port.

The entire transmission system was bench-tested with the commercial DT-DSM (AD7403) and a 2~m length of single-mode fiber. DR was measured at the DSM output as well as at the final output of the receiver for two different values of OSR. The output RF signal (at the receiver) was detected by an RF spectrum analyzer with its averaging factor set to 16 to reduce the displayed average noise level (DANL). The bandwidth of the measurements was normally set to 500~Hz with 1~Hz resolution. A 20~MHz sinusoidal signal was used as the clock for both the DSM and the DFF; this is the highest sampling frequency allowed by the DSM. The bias current used to power the laser diode was set at 25~mA; this can be further reduced to save power since loss within the fiber is low. Due to instrument limitations, the LO input was set to 120 MHz, as opposed to typical 3~T Larmor frequencies of 123-128~MHz. To test the DSM performance for different values of OSR, the RF frequency was set to either 119.8~MHz or 119.9~MHz, resulting in down-converted signal frequencies of 200~kHz ($OSR=50$) or 100~kHz ($OSR=100$), respectively.

\begin{figure}[htbp]
    \centering
    \includegraphics[width=0.75\columnwidth]{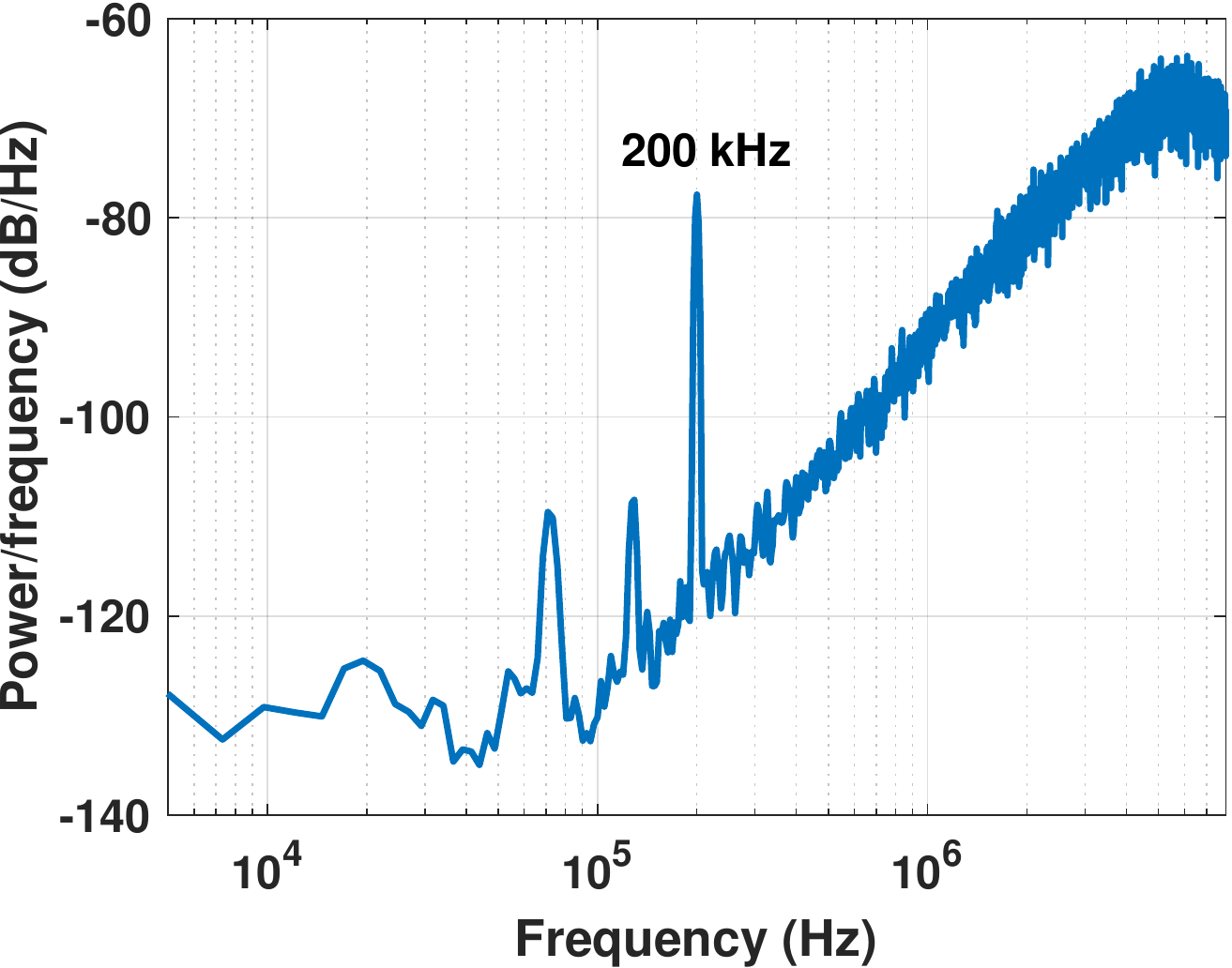}
    \caption{Typical power spectrum density (PSD) of the DSM output. The IF signal is at 200~kHz.} 
    \label{fig:5}
\end{figure}

The typical power spectral density (PSD) of the DSM output is shown in Fig.~\ref{fig:5}. In this case the RF input was set to 119.8~MHz, resulting in an IF at 200~kHz, and the encoded IF signal is visible in the spectrum of the 1-bit DSM output. Quantization noise within the IF signal bandwidth is reduced by the noise shaping feature of the DSM, as expected. However, Fig.~\ref{fig:5} also shows excess low-frequency noise that exists with or without any IF input. Possible sources of such noise are discussed in Section~\ref{sec:conclusion}.

\begin{figure*}[t]
    \centering
    \includegraphics[width=0.95\textwidth]{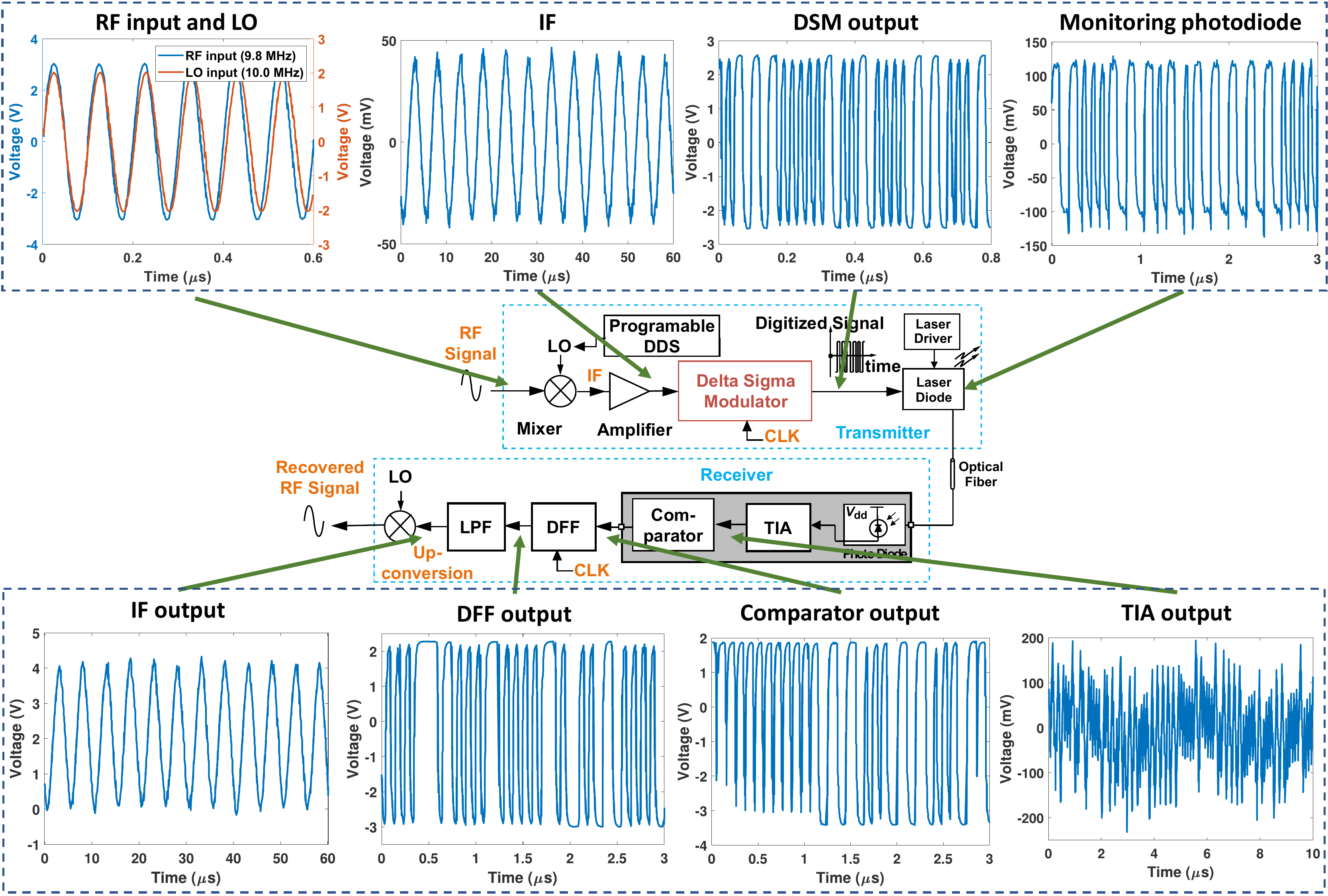}
    \caption{Measured time-domain signals at different stages of the proposed digital RF-over-fiber link.}
    \label{fig:6}
\end{figure*}

An example of the typical time domain signals measured at different stages of the link is shown in Fig.~\ref{fig:6}. The frequency mixing of two sine waves (RF and LO) results in a single IF frequency after low-pass filtering. The DSM inputs the baseband signal and outputs a digitized voltage that switches the LD on and off. The signal received by the monitoring photodiode confirms that the optical signal travelling inside the optical fiber is indeed digital. Despite the noisy signal out of the TIA (mainly caused by the limited bandwidth of the op-amp in the TIA), the final IF signal recovered by the analog low-pass filter shows a clean sinusoidal waveform at 200~kHz. The final RF signal is not displayed, because the unwanted harmonics after the up-conversion were not filtered out during our experiments. A simple passive low-pass or band-pass filter can be used to remove these harmonics.

\begin{figure}[htbp]
    \centering
    \includegraphics[width=0.80\columnwidth]{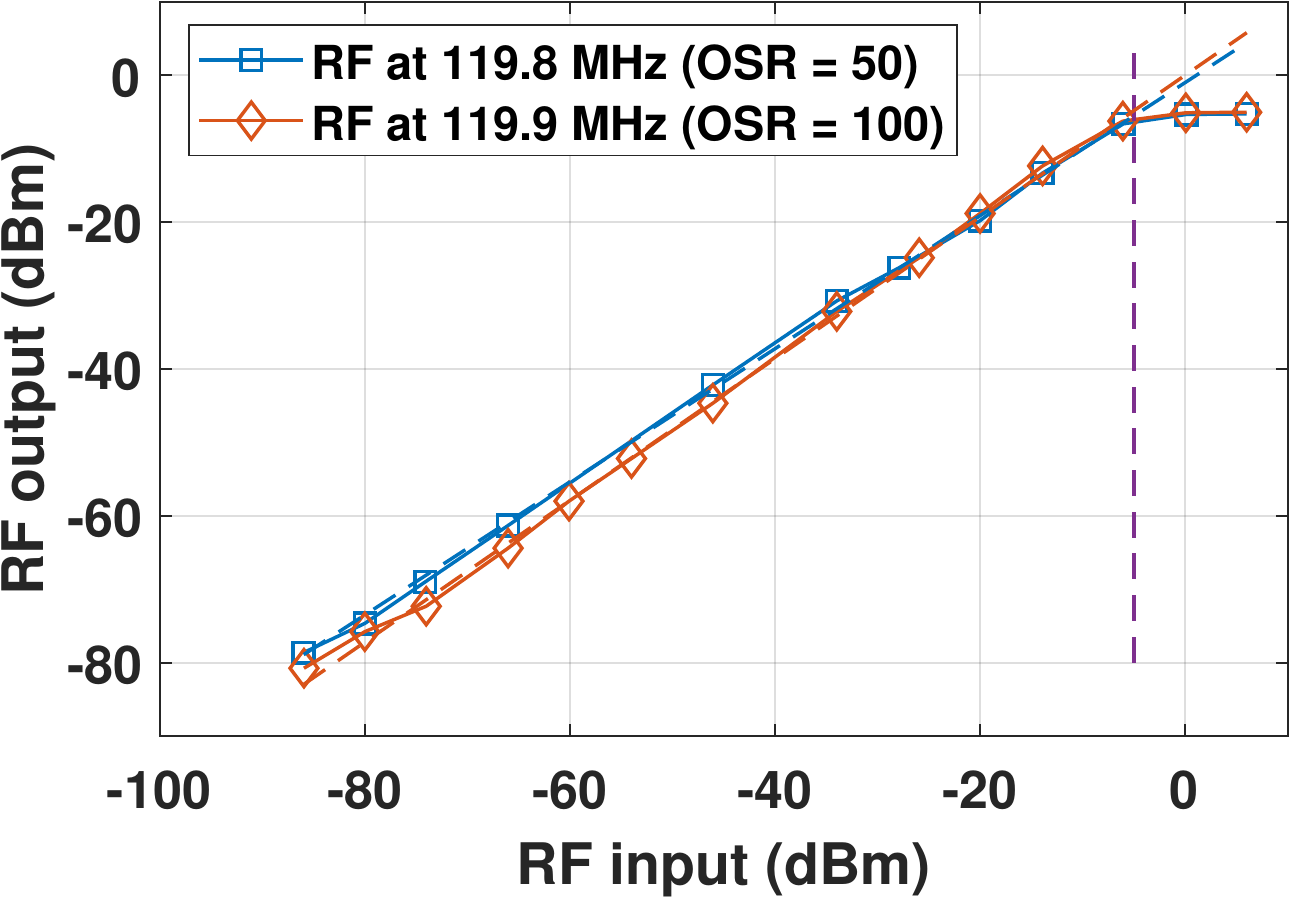}
    \caption{Measured end-to-end linearity of the link for RF inputs at different frequencies. In each case, $P_{1dB,in}$ was estimated by fitting the linear region at low input amplitudes.} 
    \label{fig:7}
\end{figure}

The final RF outputs were measured with the RF input power adjusted from $-86$~dBm to $+6$~dBm. The measured noise floors at the RF output were $-79.0$~dBm and $-80.9$~dBm for OSR of 50 and 100, respectively. The resulting RF input-output relationship is summarized in Fig.~\ref{fig:7}. A least-squares regression was performed to fit the linear region of the curves. The 1~dB compression point $P_{1dB,in}$ estimated from these fits is $-5$~dBm for both OSR values, resulting in an end-to-end DR of 81~dBm under both conditions. The fact that DR does not improve for the higher value of OSR suggests that the system's noise floor is not dominated by quantization noise (as expected for well-designed DSMs).

\subsection{Custom CT-DSM Chip}
\begin{figure}[htbp]
 	\centering
 	\includegraphics[width=0.65\columnwidth]{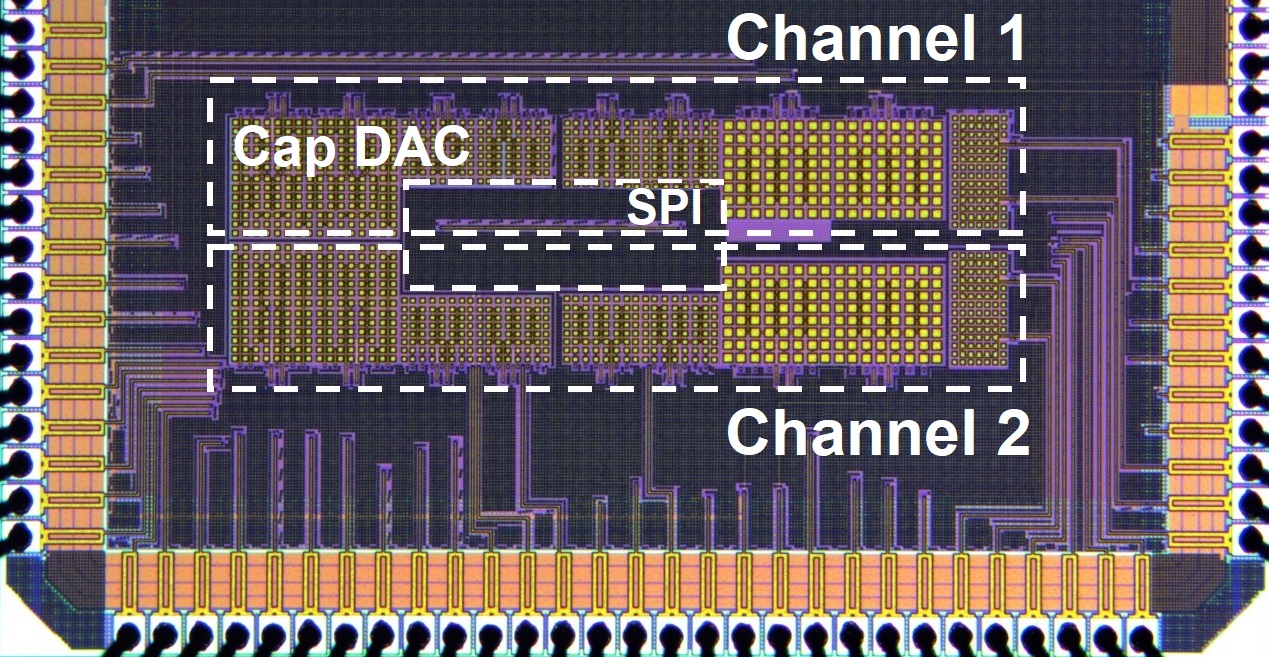}
 	\caption{Die photograph of the custom chip.}
 	\label{fig:die_photo}
 \end{figure}

Fig.~\ref{fig:die_photo} shows a die photograph of the custom CT-DSM chip, which has an active area of 2.49~mm $\times 1.51$~mm. The chip was tested on a custom PCB using a 2.2~V power supply.

\begin{figure}[htbp]
	\centering
	\includegraphics[width=0.75\columnwidth]{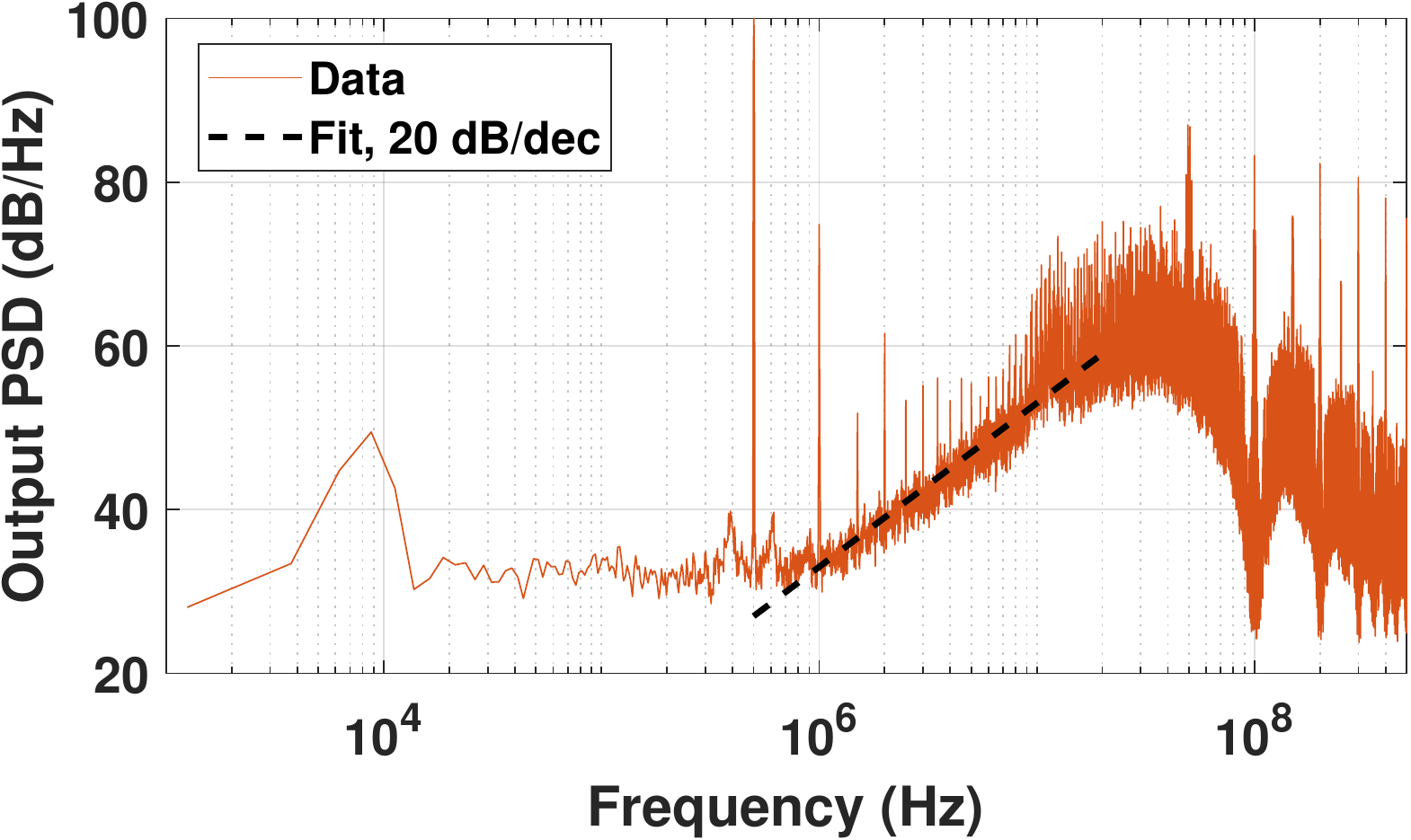}
	\caption{Measured PSD of the CT-DSM output for an input amplitude of 700~mV (which results in maximum SNR and SNDR), and a linear fit to the data. The input frequency was 500~kHz, while the clock frequency was equal to its design value of $f_s=100$~MHz.}
	\label{fig:47dBSNDR}
\end{figure}

Measurements of the output spectrum (Fig.~\ref{fig:47dBSNDR}) for single-tone inputs at 500~kHz show maximum SNR (and SNDR) at an input amplitude of 700~mV, in agreement with simulations. However, while noise shaping is visible, the estimated SNR$_{max}$ is only 47~dB due to an elevated low-frequency noise floor. In addition, the slope of the NTF is only $\sim$20~dB/decade, which is similar to that of a first-order modulator. Measurements at other input frequencies (ranging from 0-500~kHz) resulted in similar output spectra and SNR$_{max}$. Thus, the degraded NTF and SNR are likely due to three main issues: i) errors in the integrator time constants (and thus the NTF) due to process variability, ii) high levels of clock jitter due to limited slew rates at the board/chip interface, and iii) power supply noise. Next, we analyze each of these error sources in more detail.

\subsection{Analysis of On-Chip Error Sources}
\begin{figure}
    \centering
    \includegraphics[width=0.67\columnwidth]{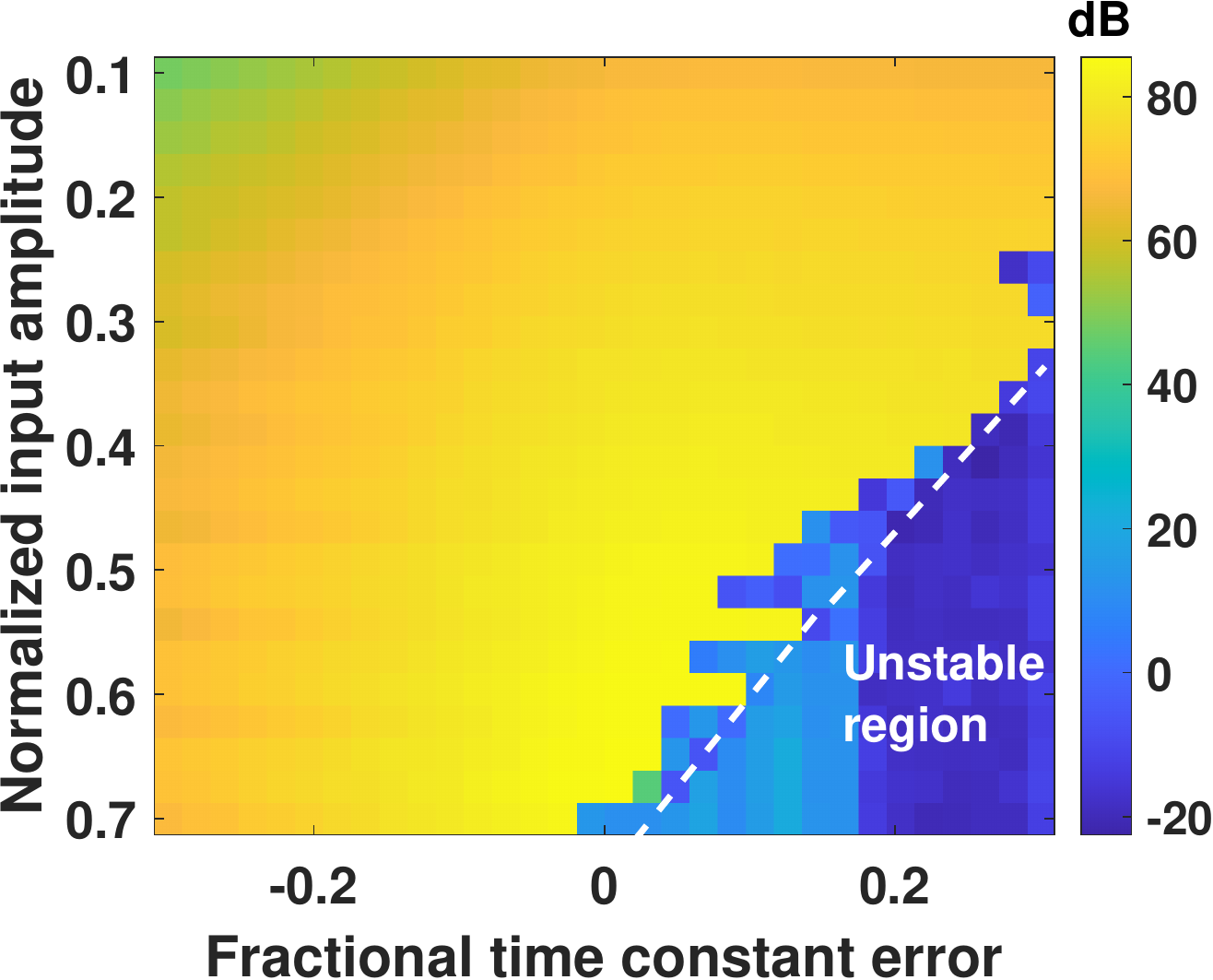}
    \caption{Simulated SQNR of the CT-DSM model in MATLAB as a function of i) amplitude of a 500~kHz sinusoidal input signal (normalized to full-scale), and ii) the fractional time constant error $\Delta k/k$ for the loop filter. The clock frequency was set to $f_{s}=100$~MHz.}
    \label{fig:dsm_mismatch}
\end{figure}

First, we consider errors in the loop filter time constants $k_{i}$, where $i=\{1,2,..,5\}$. Fig.~\ref{fig:dsm_mismatch} shows the simulated SQNR of the CT-DSM model in MATLAB as a function of the fractional time-constant error $\Delta k/k$ (assumed to be equal for all the integrators) and the input signal amplitude. Negative values of $\Delta k$ (i.e., smaller time constants than expected) result in a gradual degradation of SQNR due to changes in the positions of the NTF zeros. On the other hand, positive values (i.e., larger time constants than expected) result in the onset of instability at lower input amplitudes, which also reduces the peak SQNR. The on-chip time constants were digitally tuned over SPI to reduce these effects, but no significant improvement in SNR$_{max}$ was observed over the available tuning range (approximately $\pm 5$\%). This result suggests that time constant errors are not the dominant error source.

\begin{figure}[htbp]
    \centering
    \includegraphics[width=0.65\columnwidth]{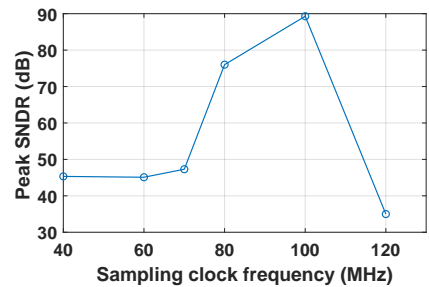}
    \caption{Simulated peak SNDR of the proposed CT-DSM circuit (not including thermal noise or clock jitter) versus $f_s$, the clock frequency.}
    \label{fig:sweep_clock}
\end{figure}

Next, we consider the effects of clock jitter. Measurements of the input clock waveform showed relatively high rise/fall times due to parasitic capacitance of the input pads and chip packaging ($\approx 0.8$~ns for 20\%-80\% V$_{pp}$, compared to the desired value of $\approx 0.1$~ns), thus resulting in significant jitter. In fact, transistor-level simulations of the CT-DSM with the observed clock rise/fall times result in i) NTFs with degraded slopes, and ii) peak SNR values that match the test results. Thus, the non-ideal clock waveform is likely to be the dominant error source. Note that this issue cannot be reduced by using a lower clock frequency $f_s$; this is because the transfer function of the continuous-time loop filter does not scale with $f_s$. As a result, the NTF (and thus the SNR) of the DSM severely degrades as $f_s$ deviates from its designed value; the effects are similar to those from time constant errors $\Delta k$ (shown in Fig.~\ref{fig:dsm_mismatch}). For example, the simulation results shown in Fig.~\ref{fig:sweep_clock} suggest that decreasing $f_s$ by 30\% from its design value (from 100~MHz to 70~MHz) degrades peak SNDR by $>40$~dB. Such sensitivity to the value of $f_s$ is a fundamental issue with CT-DSMs; it does not occur for DT-DSMs since the transfer function of a discrete-time loop filter scales with $f_s$. Thus, eliminating this issue requires redesign of the on-chip buffer that interfaces with the off-chip high-frequency clock source (typically, a crystal oscillator) to reduce rise/fall times and jitter. The existing buffer is designed for CMOS-type single-ended rail-to-rail inputs and has a capacitive input impedance (i.e., is not impedance-matched), thus resulting in limited bandwidth. Buffer performance can be significantly improved by using LVDS-type reduced-swing differential current-mode inputs and an impedance-matched interface. Alternatively, buffer design requirements can be relaxed by using an on-chip phase-locked loop (PLL) to generate a low-jitter clock from a low-frequency external reference input.

Finally, the third issue, namely power-supply noise, does not appear to be a significant error source in our system. If necessary, it can be reduced by either using passive ($LC$ or $RC$) supply filters or running the system off batteries.

\section{Discussion and Conclusion}
\label{sec:conclusion}
This work investigated the feasibility of using DSM-based digital RF-over-fiber links to transmit and receive MRI signals from array coils. The biggest challenge for such non-coaxial transmission systems is the large DR of typical MRI signals. Most imaging applications with typical 3T~MRI scanners require a DR of at least 80~dB over a bandwidth of 1~MHz. An initial bench test of an RF-over-fiber link was conducted using a commercial second-order DSM with a maximum sampling frequency of 20 MHz, and sufficient end-to-end DR of 81~dB was demonstrated for IF signals at 200~kHz (\(OSR=50\)). However, because of excess low-frequency noise, the DR remained at 81~dB for increased OSR (\(OSR=100\)).

To identify the source of the low-frequency noise, the DR tests were repeated by replacing the DC power supply with dry-cell batteries. However, no significant reduction in the noise floor was observed at either the DSM output or the final receiver output. Therefore, power supply noise and/or ripple is not a major source. Also, the fact that the noise is already present at the DSM output implies that LD intensity fluctuations are unlikely to be important either. Thus, the low-frequency noise is most likely caused by the down-converted phase noise of the LO. Once this source is suppressed (e.g., by using a lower-noise LO), the optical link should provide a DR close to the AD7403's peak SNR of 88~dB at an OSR of 100. Replacing this part with the custom high-order high-frequency CT-DSM, which is expected to provide a peak SNR of $\sim$85 dB at an OSR of 50, would allow the link to achieve $>80$~dB DR over a \(\pm 500\) kHz bandwidth, which would be sufficient for most practical MRI applications.

In addition, the optical link has relatively low power consumption compared to other non-coaxial transmission systems. During our bench tests, the transmitter and the receiver were both powered by a 5~V supply and the total power consumption was about 100~mW. However, this can be largely reduced by adjusting the laser driving circuit. The laser diode is typically driven by a bias current of 30~mA with an output light power of a 2.5~mW. However, the resistor that sets the bias current alone dissipates $P = I_{bias}\cdot V_{ref}\approx 75$~mW in the present design. Both the bias current and the value of $V_{ref}$ can be decreased to reduce this power dissipation. Because the transmitter relies on an on-and-off (i.e., OOK) modulation scheme, high bias current is unnecessary. The bias current only needs to cross the threshold current of the laser diode, which is typically 9~mA. If \(I_{bias}\) and \(V_{ref}\) are both halved, the laser driver can easily consume less than 20~mW while maintaining the ability to power the laser diode. The custom CT-DSM chip has a power consumption of approximately 20~mW, so the entire high-bandwidth transmitter can consume $<50$~mW. The low power consumption of this system would allow the use of a non-magnetic battery as the power supply, which eliminates the undesired coaxial cables and makes the system more compatible with existing MRI coil arrays.

In conclusion, DSM-based digital RF-over-fiber links are promising for replacing coaxial cables as MRI interconnects. Our prototype link used a commercial second-order DSM to achieve $>80$~dB end-to-end DR at high OSR, and this number is expected to further improve by $\sim$7~dB if a more stable LO is used to reduce low-frequency noise. Future advances in the design of the customized high-speed and high-order CT-DSM chip can be readily incorporated into the new RF-over-fiber system presented herein for transmitting full-bandwidth MRI signals. For this purpose, the off-the-shelf mixer, amplifier, and DSM blocks used in the basic link (see Fig.~\ref{fig:1}) will be replaced by the custom chip. In addition, two matched copies of the laser driver, laser diode, optical fiber, and receiver will be implemented to transmit and decode the I and Q signal components in parallel (see Fig.~\ref{fig:improved_link}). Alternatively, wavelength-division multiplexing (WDM) can be used to transmit both components on a single fiber, as mentioned earlier.  

\begin{acknowledgments}
The authors wish to thank Quality Electrodynamics, LLC for funding, and David Ariando for his assistance with setting up and programming the DDS. Support for this work also comes from the Ohio Third Frontier program.
\end{acknowledgments}

\section*{Data Availability}
The data that support the findings of this study are available from the corresponding author
upon reasonable request.

\nocite{*}
\bibliography{Paper}

\end{document}